\documentclass[nofootinbib,aps,a4paper,letterpaper,superscriptaddress,
twocolumn,showpacs,times,eqsecnum]{revtex4}
\usepackage{amsmath}
\usepackage{amsfonts}
\usepackage{booktabs}
\usepackage{multirow}
\usepackage{siunitx} 
\usepackage{adjustbox}
\pdfoutput=1
\usepackage{graphicx}
\usepackage{color}
\usepackage{braket}
\usepackage{dcolumn}
\usepackage{bm,url}
\usepackage[linktocpage]{hyperref}
\usepackage{subfigure}
\usepackage{amsfonts}
\usepackage[usenames,dvipsnames,svgnames]{xcolor}  
\usepackage{hyperref}   
\definecolor{oxfordblue}{rgb}{0.0, 0.13, 0.28}
\definecolor{burgundy}{rgb}{0.5, 0.0, 0.13}
\definecolor{darkolivegreen}{rgb}{0.33, 0.42, 0.18}
\definecolor{darkblue}{rgb}{0,0,0.5}
\definecolor{richcarmine}{rgb}{0.84, 0.0, 0.25}
\definecolor{darkblue}{rgb}{0,0,0.5}
\definecolor{bluer}{rgb}{0.00,0.50,0.75}{}
\hypersetup{colorlinks=true, citecolor=red, linkcolor=blue,
 urlcolor = magenta, filecolor=magenta}

\begin{document}
 
 \newcommand\be{\begin{equation}}
  \newcommand\ee{\end{equation}}
 \newcommand\bea{\begin{eqnarray}}
  \newcommand\eea{\end{eqnarray}}
 \newcommand\bseq{\begin{subequations}} 
  \newcommand\eseq{\end{subequations}}
 \newcommand\bcas{\begin{cases}}
  \newcommand\ecas{\end{cases}}
 \newcommand{\p}{\partial}
 \newcommand{\f}{\frac}

 \title{Exact reconstruction of running vacuum cosmology in torsion and nonmetricity
gravity}

 \author{Fotios K. Anagnostopoulos }
\affiliation{Department of Informatics and Telecommunications, University of
Peloponnese, Karaiskaki 70, 22100, Tripoli, Greece}

\author{Spyros Basilakos}
\affiliation{National Observatory of Athens, Lofos Nymfon, 11852 Athens, Greece}
\affiliation{Academy of Athens, Research Center for Astronomy and Applied
Mathematics, Soranou Efesiou 4, 11527, Athens, Greece}
\affiliation{School of Sciences, European University Cyprus, Diogenes Street,
Engomi, 1516 Nicosia, Cyprus}

 \author{ Eftychios E. Chalkiadakis}
  \affiliation{Department of Physics, School of Sciences, University of
Thessaly,
3rd Km Old National Road, 35100, Lamia, Greece}
\affiliation{Institute for Astronomy, Astrophysics, Space Applications and
Remote Sensing, National Observatory of Athens, 15236 Penteli, Greece}

  \author{Stefanos Pigkas-Balanikas}
    \affiliation{Department of Physics, University of Patras, 26504 Patras, Greece}
\affiliation{Institute for Astronomy, Astrophysics, Space Applications and
Remote Sensing, National Observatory of Athens, 15236 Penteli, Greece}

  \author{ Panagiotis G. Stavros}
  \affiliation{Department of Physics, School of Sciences, University of
Thessaly,
3rd Km Old National Road, 35100, Lamia, Greece}
\affiliation{Institute for Astronomy, Astrophysics, Space Applications and
Remote Sensing, National Observatory of Athens, 15236 Penteli, Greece}

\author{Emmanuel N. Saridakis}
 \affiliation{Institute for Astronomy, Astrophysics, Space Applications and
Remote Sensing, National Observatory of Athens, 15236 Penteli, Greece}
 \affiliation{Departamento de Matem\'{a}ticas, Universidad Cat\'{o}lica del
  Norte, Avda. Angamos 0610, Casilla 1280, Antofagasta, Chile}
 \affiliation{CAS Key Laboratory for Research in Galaxies and Cosmology,
School
  of Astronomy and Space Science,
  University of Science and Technology of China, Hefei 230026, China}

\begin{abstract}

Running vacuum cosmology provides a dynamical description
of vacuum energy while retaining the strict equation of
state $w_{\rm DE}=-1$. However, whether its complete
cosmological dynamics can emerge from a gravitational
Lagrangian remains an important theoretical question.
In this work, we investigate this possibility within
torsion and nonmetricity gravity, considering minimally
coupled, nonminimal matter-coupled, and trace-coupled
theories. We first reconstruct analytically the running
vacuum energy density  in
$f(T)$ and $f(Q)$ gravity, showing that energy density matching
does not generally imply identical equations of state
or expansion histories. We quantify these differences
using Pantheon+, cosmic chronometers, and DESI DR2 BAO
data. However,   considering  trace-coupled
$f(T,\mathcal{T})$ and $f(Q,\mathcal{T})$ gravity we extract
explicit gravitational Lagrangians  that exactly reproduce the
complete running vacuum dynamics at the background
level, namely  both Friedmann
equations, the strict vacuum equation of state,
the matter-vacuum interaction, and the resulting
expansion history. Furthermore, we identify regular
solutions approaching a future de Sitter attractor,
with positive matter and vacuum densities. Our results establish an explicit
gravitational realization of running vacuum cosmology,
extending beyond the mere matching of its effective
vacuum energy density.

\end{abstract}

 \maketitle

\section{Introduction}
\label{sec:introduction}

Despite its remarkable success, the standard $\Lambda$CDM
model faces important theoretical and observational challenges.
The nature of dark matter and dark energy remains unknown,
while the cosmological constant problem persists
\cite{Weinberg:1988cp,Bertone:2004pz}.
Moreover, various tensions in cosmological measurements
have intensified interest in alternative scenarios
\cite{Abdalla:2022yfr,CosmoVerseNetwork:2025alb}.
These developments have motivated three main approaches, namely
retaining general relativity while introducing additional
fields or effective fluids, such as quintessence, k-essence,
and interacting dark energy
\cite{Copeland:2006wr,Cai:2009zp}, using entropic considerations to alter the
cosmological behavior \cite{Saridakis:2026vxo}, or   modifying the gravitational
sector itself to account
for cosmic acceleration
\cite{CANTATA:2021asi,Clifton:2011jh,
Capozziello:2011et,Nojiri:2017ncd}.

Within the third direction, classes of modified gravity theories
can be constructed using curvature, torsion, or
nonmetricity as the fundamental geometrical quantity.
In particular, torsion and nonmetricity
formulations have attracted considerable interest,
leading to
torsional $f(T)$, $f(T,T_G)$, and $f(T,B)$ theories
\cite{Cai:2015emx,Ferraro:2006jd,Linder:2010py,
Chen:2010va,Tamanini:2012hg,Kofinas:2014owa,
Kofinas:2014daa,Bahamonde:2015zma,Bahamonde:2016grb},
as well as  nonmetricity-based $f(Q)$ and $f(Q,C)$ gravity
\cite{BeltranJimenez:2017tkd,BeltranJimenez:2019tme,
Anagnostopoulos:2021ydo,Heisenberg:2023lru,
De:2023xua}. Furthermore, one can extend these scenarios by introducing nonminimal
interactions of geometry with the matter sector, giving rise to
nonminimal torsion-matter-coupling theories \cite{Harko:2014sja} and
$f(T,\mathcal{T})$ gravity (with $\mathcal{T}$ the trace of the energy-momentum
tensor) \cite{Harko:2014aja}, as well as to nonminimal nonmetricity-matter-coupling
theories \cite{Harko:2018gxr} and
 $f(Q,\mathcal{T})$ gravity \cite{Xu:2019sbp}.   Such couplings enrich the
cosmological dynamics
and may induce an effective exchange of energy between
matter and the geometrical sector \cite{Junior:2015bva, Carloni:2015lsa,
Saez-Gomez:2016wxb, Farrugia:2016pjh, Pace:2017aon, Pace:2017dpu, Bahamonde:2017ifa,
Xu:2020yeg, Ghosh:2020rau, Arora:2020tuk, Gonzalez-Espinoza:2021mwr,
Duchaniya:2023vhn, Mishra:2023khd,Loo:2023oxk, Mishra:2023onl,
Samaddar:2024qno,Anagnostopoulos:2026vow}.
For a spatially flat Friedmann-Lema\^itre-Robertson-Walker (FLRW) spacetime the
torsion scalar is $T=6H^2$, while in the same spacetime and with the basic symmetric
teleparallel connection (Connection I) the nonmetricity scalar is
$Q=6H^2$.
This implies that the corresponding torsional and
nonmetricity theories share the same background
cosmological equations. However, their dynamics
need not coincide at the perturbative level,
for other spacetime geometries, or when considering
different connection branches (such as Connection II and Connection III)
\cite{Hohmann:2020zre,Hohmann:2021ast,DAmbrosio:2021pnd,
Paliathanasis:2023nkb,Paliathanasis:2015aos,Dimakis:2016mip,
Heisenberg:2022mbo,Shi:2023kvu,
Wu:2024vcr,GhulamMurtaza:2026njh}.

On the other hand, an interesting alternative to
conventional dynamical dark energy is provided by
running vacuum cosmology
\cite{Shapiro:2000dz,Basilakos:2009wi,Basilakos:2013vya,Sola:2016jky}.
In this framework, the vacuum energy density is
allowed to evolve with the cosmological expansion,
while retaining the characteristic vacuum equation
of state $p_\Lambda=-\rho_\Lambda$.
The motivation originates from quantum field theory
in curved spacetime, where the renormalization of
vacuum fluctuations can generate contributions to
the effective vacuum energy that depend on the
curvature and the cosmological expansion rate
\cite{Shapiro:2000dz,Moreno-Pulido:2020anb,Moreno-Pulido:2022phq}.
Consequently, running vacuum scenarios provide
a framework for describing different accelerated
epochs through the evolution of the vacuum sector,
without necessarily introducing additional
fundamental scalar fields, and  their implications for
the expansion history, structure formation, and
cosmological tensions have been extensively investigated
  \cite{Basilakos:2012ra, Bouhmadi-Lopez:2013mji,
Basilakos:2015yoa, Lima:2015mca, Basilakos:2015vra, Fritzsch:2016ewd,
Oikonomou:2016pnq, Kohri:2016lsj, Geng:2017apd,
 Basilakos:2019zsf, Basilakos:2019acj, Panotopoulos:2020kpo,
Basilakos:2020qmu, Mavromatos:2020crd, Mavromatos:2020kzj,
Mavromatos:2021urx, Rezaei:2021qwd, Jawad:2021bas, Cruz:2023dzn,
Gomez-Valent:2023hov}.

Since in running vacuum cosmology the vacuum energy evolves while retaining
the strict equation of state $w_\Lambda=-1$, for constant Newton's
gravitational coupling
the Bianchi identities   imply an energy exchange
between vacuum and matter, leading to a modified
matter conservation equation. Therefore, running vacuum cosmology constitutes
an intrinsically interacting dark energy scenario, and thus shares the successes of
interacting dark energy
 \cite{Amendola:1999er, Billyard:2000bh, Amendola:2000uh, Zimdahl:2001ar, Cai:2004dk,
delCampo:2008sr, Cai:2009ht, Faraoni:2014vra,Koivisto:2015qua,Wang:2016lxa,
Kumar:2019wfs, Paliathanasis:2024jxo, Khoury:2025txd, Silva:2025hxw,
Bhattacharjee:2025xeb, Paliathanasis:2025xxm, Yang:2026vkc}.

 A long-standing  theoretical question concerns the
gravitational Lagrangian realization of running
vacuum cosmology. Although quantum field theory
in curved spacetime motivates a dynamical vacuum
density,
prescribing $\rho_\Lambda(H)$ and
$p_\Lambda=-\rho_\Lambda$ does not automatically
provide a gravitational action reproducing the
complete cosmological evolution. In particular, matching
the effective vacuum energy density is only partially sufficient unless
the field equations also reproduce the strict vacuum
equation of state and the associated matter-vacuum
interaction. This raises the question of whether
running vacuum cosmology can emerge directly from
modified gravitational dynamics.

In the present work, we investigate the gravitational
realization of running vacuum cosmology within torsion
and nonmetricity theories. We first reconstruct the
running vacuum energy density
within minimally coupled $f(T)$ and $f(Q)$ gravity.
Although the energy density reconstruction is exact and the cosmological evolutions
of these models and running vacuum  are quite similar, the
resulting equation of state and expansion history
are   different,  which is expected since simple
$f(T)$ and $f(Q)$ gravity are not intrinsically interacting.
In order to investigate these features we confront the scenarios
using Pantheon+ supernovae,
cosmic chronometers, and DESI DR2 BAO data, and we apply information criteria such
as AIC and BIC. Additionally, we study the reconstruction of running vacuum
cosmology within the    nonminimal geometry-matter couplings, showing  that the
simultaneous satisfaction of all requirements recovers standard $\Lambda$CDM
cosmology.

We then demonstrate that trace-coupled
$f(T,\mathcal{T})$ and $f(Q,\mathcal{T})$ gravity
can fully  reproduce the complete running vacuum cosmology. In particular,
our reconstruction   reproduces the exact vacuum energy density alongside the strict
relation $w_{\rm DE}=-1$, the matter-vacuum interaction,
and the exact running vacuum  expansion history, establishing
a full gravitational realization.

The manuscript is organized as follows.
In Sec. \ref{sec:rvm} we review running vacuum cosmology
and its matter-vacuum interaction.
In Sec. \ref{sec:modified_gravity} we present the torsional and
nonmetricity theories considered in our analysis.
In Sec. \ref{sec:reconstruction} we reconstruct the running vacuum
density in minimally and nonminimally coupled
models and investigate the observational
implications. In Sec. \ref{sec:exact_rvm} we derive the conditions
for exact running vacuum equivalence and obtain
explicit linear and nonlinear trace-coupled
reconstructions. Finally, in Sec. \ref{sec:conclusions} we
summarize our conclusions.

 \section{Running vacuum cosmology}
\label{sec:rvm}

In this section, we briefly review the running vacuum
framework that will serve as the reference for our
gravitational reconstructions. We first present the
cosmological equations and expansion history, and we then examine the matter-vacuum
interaction required by the conservation equations.

\subsection{Running vacuum energy and cosmological evolution}
\label{subsec:rvm_background}

The running vacuum scenario constitutes an interesting
alternative to the standard cosmological constant paradigm,
in which the vacuum energy density is allowed to evolve
during the cosmic expansion, while maintaining the
characteristic vacuum equation of state
\begin{equation}
p_{\Lambda}=-\rho_{\Lambda}.
\label{eq:rvm_eos}
\end{equation}
Hence, contrary to conventional dynamical dark energy models,
where the evolution is typically associated with a varying
equation-of-state parameter, in running vacuum cosmology
the dynamical behavior is entirely encoded in the time
dependence of the vacuum energy density.

In order to investigate the cosmological evolution, we
consider a spatially flat Friedmann-Lema\^itre-Robertson-Walker
(FLRW) geometry, described by the line element
\begin{equation}
ds^2=-dt^2+a^2(t)\delta_{ij}dx^i dx^j,
\label{eq:rvm_metric}
\end{equation}
where $a(t)$ is the scale factor and $H=\dot a/a$ is
the Hubble parameter, with dots denoting derivatives
with respect to cosmic time.

The running vacuum framework can be motivated by quantum
field theory in curved spacetime, where the vacuum energy
density may acquire contributions depending on the Hubble
parameter and its derivatives
\cite{Sola:2016jky}.
In particular, the basic form of the vacuum energy
density is
\begin{equation}
\rho_{\Lambda}(H)
=
\frac{3}{8\pi G}
\left(
c_0+\nu H^2+\gamma H^4
\right),
\label{eq:rvm_density}
\end{equation}
where $c_0$, $\nu$, and $\gamma$ are constants, and
$G$ is Newton's gravitational constant.
The constant term $c_0$ corresponds to the standard
cosmological constant contribution, while the terms
proportional to $H^2$ and $H^4$ introduce the dynamical
vacuum behavior. In particular, the quadratic contribution
is relevant for the late-time cosmological evolution,
whereas the quartic term can become important at
large values of the Hubble parameter, and may play
a significant role in the early Universe.
For $\nu=\gamma=0$, the standard cosmological
constant scenario is recovered.

In the presence of pressureless matter
and running vacuum energy, the Friedmann equations
take the form
\begin{equation}
3H^2=8\pi G
\left(\rho_m+ \rho_{\Lambda}\right),
\label{eq:rvm_friedmann1}
\end{equation}
and
\begin{equation}
2\dot H+3H^2
=-8\pi G\left(p_m-\rho_{\Lambda}\right),
\label{eq:rvm_friedmann2}
\end{equation}
where $\rho_m$, $p_m$   denote the matter  energy density and pressure,
respectively (since our main interest lies in the correspondence
between running vacuum cosmology and modified
gravity at the level of the late-time background
evolution,   we neglect the
radiation contribution). In this case,
substituting Eq.~(\ref{eq:rvm_density}) into
Eq.~(\ref{eq:rvm_friedmann1}), we obtain
\begin{equation}
\rho_m
=
\frac{3}{8\pi G}
\left[
(1-\nu)H^2-c_0-\gamma H^4
\right].
\label{eq:rvm_matter}
\end{equation}
Thus, focusing on the case of dust matter, i.e. $p_m=0$,  the cosmological
evolution is governed by
\begin{equation}
\dot H
=
-\frac{3}{2}
\left[
(1-\nu)H^2-c_0-\gamma H^4
\right].
\label{eq:rvm_evolution}
\end{equation}
The above equation determines the background
expansion history for the considered running
vacuum model.

\subsection{Matter-vacuum interaction and conservation equations}
\label{subsec:rvm_interaction}

An essential feature of running vacuum cosmology is that,
although the vacuum energy density evolves during the
cosmological expansion, its equation of state remains
identically equal to $-1$. This has important consequences
for the conservation equations. In particular, for constant
Newton's gravitational constant, the time dependence of
the vacuum energy density necessarily implies an energy
exchange between the vacuum and the matter sector.

Indeed, the Bianchi identities imply the conservation of
the total energy-momentum tensor, leading to
\begin{equation}
\dot{\rho}_m+\dot{\rho}_{\Lambda}
+3H(\rho_m+ \rho_{\Lambda}+p_{\Lambda})=0.
\label{eq:rvm_total_conservation}
\end{equation}
Taking into account the vacuum equation of state
$p_{\Lambda}=-\rho_{\Lambda}$, and considering pressureless
matter, the above equation reduces to
\begin{equation}
\dot{\rho}_m+3H\rho_m=-\dot{\rho}_{\Lambda}.
\label{eq:rvm_interaction}
\end{equation}
Hence, if the matter sector were separately conserved,
namely if $\dot{\rho}_m+3H\rho_m=0$, one would necessarily
obtain $\dot{\rho}_{\Lambda}=0$, corresponding to a constant
vacuum energy density. Therefore, a genuinely running
vacuum with $\dot{\rho}_{\Lambda}\neq 0$ requires an
interaction between matter and vacuum energy.

The corresponding conservation equations can be written
as
\begin{align}
\dot{\rho}_m+3H\rho_m &= \mathcal{I},
\label{eq:rvm_matter_conservation}\\
\dot{\rho}_{\Lambda} &= -\mathcal{I},
\label{eq:rvm_vacuum_conservation}
\end{align}
where $\mathcal{I}$ is the interaction function describing
the energy exchange between the two sectors. Positive
values of $\mathcal{I}$ correspond to energy transfer from
vacuum to matter, while negative values describe the
opposite process.

For the running vacuum density given by
Eq.~(\ref{eq:rvm_density}), the interaction function
becomes
\begin{equation}
\mathcal{I}
=-\frac{3}{8\pi G}
\left(2\nu H+4\gamma H^3\right)\dot H,
\label{eq:rvm_interaction_general}
\end{equation}
while  using (\ref{eq:rvm_evolution})
we finally obtain
\begin{equation}
\mathcal{I}
=3H\left(\nu+2\gamma H^2\right)\rho_m.
\label{eq:rvm_interaction_explicit}
\end{equation}
Thus, the interaction is determined by the running
vacuum parameters and the cosmological evolution.
In particular, for $\gamma=0$ the interaction reduces
to $\mathcal{I}=3\nu H\rho_m$, while for
$\nu=\gamma=0$ it vanishes, recovering the standard
$\Lambda$CDM scenario.

The above discussion reveals that the interaction
between matter and vacuum energy is not an additional
assumption, but a necessary consequence of combining
a time-dependent vacuum density with the strict vacuum
equation of state and constant gravitational coupling.
This observation provides a natural motivation for
investigating gravitational theories with nonminimal
geometry-matter couplings, where an effective energy
exchange between matter and the geometrical dark energy
sector can arise directly from the gravitational field
equations. In the following sections, we examine whether
such theories can reproduce the running vacuum energy
density, and under which conditions the exact vacuum
equation of state can also be recovered.

\section{Torsion and nonmetricity theories of gravity}
\label{sec:modified_gravity}

In this section, we briefly present the torsional and
nonmetricity gravitational theories employed in our
analysis. As we mentioned in the Introduction, in this work we focus on
nonmetricity theories with the basic symmetric teleparallel connection (Connection
I), which for spatially flat FLRW cosmology coincide with their torsional
counterparts at the background
level  \cite{Hohmann:2020zre,Hohmann:2021ast,DAmbrosio:2021pnd,
Paliathanasis:2023nkb,Paliathanasis:2015aos,Dimakis:2016mip,
Heisenberg:2022mbo,Shi:2023kvu,
Wu:2024vcr,GhulamMurtaza:2026njh},
allowing for a unified treatment of the two
geometrical formulations. Nevertheless,
this correspondence requires consistent
matter prescriptions and does not imply
complete dynamical equivalence.
In particular, the trace-coupled theories possess
identical homogeneous cosmological equations when
a common perfect-fluid variation prescription is
adopted. Nevertheless, cosmological perturbations
or different gravitational connections may lead
to distinct dynamics.

\subsection{$f(T)$ and $f(Q)$ gravity}
\label{subsec:fT_fQ}

In a general metric-affine geometry, the gravitational
connection $\Gamma^{\lambda}{}_{\mu\nu}$ is independent of
the metric and can be decomposed as
\begin{equation}
\Gamma^{\lambda}{}_{\mu\nu}
=
\left\{^{\lambda}_{\mu\nu}\right\}
+K^{\lambda}{}_{\mu\nu}
+L^{\lambda}{}_{\mu\nu},
\label{eq:connection_decomposition}
\end{equation}
where $\left\{^{\lambda}_{\mu\nu}\right\}$ is the
Levi-Civita connection, while
$K^{\lambda}{}_{\mu\nu}$ and
$L^{\lambda}{}_{\mu\nu}$ are the contorsion and
disformation tensors, respectively.
The latter are constructed from the torsion and
nonmetricity tensors, defined as
\begin{equation}
T^{\lambda}{}_{\mu\nu}
\equiv 2\Gamma^{\lambda}{}_{[\mu\nu]},
\qquad
Q_{\alpha\mu\nu}
\equiv \nabla_{\alpha}g_{\mu\nu}.
\label{eq:torsion_nonmetricity_definitions}
\end{equation}

In General Relativity, the connection is chosen to
be the Levi-Civita one, and hence both torsion and
nonmetricity vanish. Alternatively, in teleparallel
gravity one employs a curvature-free and
metric-compatible connection, namely
$
R^{\lambda}{}_{\rho\mu\nu}=0$ and
$Q_{\alpha\mu\nu}=0$,
while the torsion tensor is in general nonvanishing.
In the Weitzenb\"ock gauge, the corresponding
connection takes the form
$
\Gamma^{\lambda}{}_{\mu\nu}
=e_A{}^{\lambda}\partial_{\mu}e^A{}_{\nu}$,
where $e^A{}_{\mu}$ denotes the tetrad field, and the gravitational interaction is
  described
through the torsion scalar  $
T = \frac{1}{4} T^{\rho\mu\nu} T_{\rho\mu\nu}
+ \frac{1}{2} T^{\rho\mu\nu} T_{\nu\mu\rho}
- T_\rho T^\rho$.

On the other hand, in symmetric teleparallel gravity
one considers a connection that is both curvature-free
and torsion-free, namely
$
R^{\lambda}{}_{\rho\mu\nu}=0$ and
$T^{\lambda}{}_{\mu\nu}=0$,
while nonmetricity is in general nonvanishing.
The corresponding connection is locally of pure-gauge
form, i.e.
$\Gamma^{\lambda}{}_{\mu\nu}
=
\frac{\partial x^{\lambda}}{\partial \xi^{\alpha}}
\partial_{\mu}\partial_{\nu}\xi^{\alpha}$,
where $\xi^{\alpha}$ are four functions specifying
the connection. In particular, choosing
$\xi^{\alpha}=x^{\alpha}$ leads to the coincident
gauge, in which
$\Gamma^{\lambda}{}_{\mu\nu}=0$.
In this formulation, the gravitational interaction
is described through the nonmetricity scalar $
Q = \frac{1}{4}Q_{\alpha\mu\nu}Q^{\alpha\mu\nu}
-\frac{1}{2}Q_{\alpha\mu\nu}Q^{\mu\alpha\nu}
-\frac{1}{4}Q_\alpha Q^\alpha
+\frac{1}{2}Q_\alpha\tilde Q^\alpha $, with $
Q_\alpha=Q_{\alpha\mu}{}^\mu$,
$\tilde Q_\alpha=Q^\mu{}_{\alpha\mu}$.

Although the teleparallel and symmetric teleparallel
formulations are dynamically equivalent to General
Relativity, their nonlinear extensions lead to distinct
modified gravitational theories.
In the torsional formulation, the gravitational action
is generalized to \cite{Cai:2015emx,Ferraro:2006jd,Linder:2010py}
\begin{equation}
S_T=\frac{1}{16\pi G}\int d^4x\,e
\left[-T+f(T)\right]+S_m,
\label{eq:fT_action}
\end{equation}
where $T$ is the torsion scalar, $e$ is the tetrad
determinant, and $S_m$ is the matter action.
On the other hand, in the symmetric teleparallel
formulation, where gravity is attributed to nonmetricity,
the corresponding action can be written as
\begin{equation}
S_Q=\frac{1}{16\pi G}\int d^4x\,\sqrt{-g}
\left[-Q+\tilde f(Q)\right]+S_m,
\label{eq:fQ_action}
\end{equation}
where $Q$ is the nonmetricity scalar.
Note that for the spatially flat FLRW geometry, and adopting
the above geometrical conventions and mostly-plus metric signature, the
corresponding
geometrical scalars become
\begin{equation}
T=6H^2,\qquad Q=6H^2.
\label{eq:T_Q_FLRW}
\end{equation}

The Friedmann equations of $f(T)$ gravity
are
\begin{align}
3H^2 &= 8\pi G\rho_m-\frac{f}{2}+Tf_T,
\label{eq:fT_friedmann1}\\
\dot H &=
-\frac{4\pi G(\rho_m+p_m)}
{1-f_T-2Tf_{TT}},
\label{eq:fT_friedmann2}
\end{align}
where subscripts denote derivatives with respect to
the torsion scalar.
Observing the form of the above equations, we can
identify an effective dark energy sector of
gravitational origin. In particular, for pressureless
matter, the effective dark energy density and
pressure are given by
\begin{align}
\rho_{\rm DE}^{(T)}
&=
\frac{1}{16\pi G}
\left(-f+2Tf_T\right),
\label{eq:fT_rhoDE}\\
p_{\rm DE}^{(T)}
&=
\frac{1}{16\pi G}
\frac{f-Tf_T+2T^2f_{TT}}
{1-f_T-2Tf_{TT}},
\label{eq:fT_pDE}
\end{align}
and hence the corresponding equation-of-state
parameter becomes
\begin{equation}
w_{\rm DE}^{(T)}
=
-\frac{f/T-f_T+2Tf_{TT}}
{(1-f_T-2Tf_{TT})(f/T-2f_T)}.
\label{eq:fT_wDE}
\end{equation}

Similarly, the background equations of $f(Q)$ gravity,
for the standard symmetric teleparallel FLRW connection,
read
\begin{align}
3H^2 &= 8\pi G\rho_m-\frac{\tilde f}{2}
+Q\tilde f_Q,
\label{eq:fQ_friedmann1}\\
\dot H &=
-\frac{4\pi G(\rho_m+p_m)}
{1-\tilde f_Q-2Q\tilde f_{QQ}}.
\label{eq:fQ_friedmann2}
\end{align}
In this case, the effective dark energy density
and pressure for pressureless matter are
\begin{align}
\rho_{\rm DE}^{(Q)}
&=
\frac{1}{16\pi G}
\left(-\tilde f+2Q\tilde f_Q\right),
\label{eq:fQ_rhoDE}\\
p_{\rm DE}^{(Q)}
&=
\frac{1}{16\pi G}
\frac{\tilde f-Q\tilde f_Q+2Q^2\tilde f_{QQ}}
{1-\tilde f_Q-2Q\tilde f_{QQ}},
\label{eq:fQ_pDE}
\end{align}
while the corresponding equation-of-state
parameter becomes
\begin{equation}
w_{\rm DE}^{(Q)}
=
-\frac{\tilde f/Q-\tilde f_Q+2Q\tilde f_{QQ}}
{(1-\tilde f_Q-2Q\tilde f_{QQ})
(\tilde f/Q-2\tilde f_Q)}.
\label{eq:fQ_wDE}
\end{equation}

It is straightforward to verify that the two sets of
cosmological equations coincide under the identification
$\tilde f(Q)=f(Q)$. Hence, despite their different
geometrical foundations, the two theories admit identical
homogeneous background evolutions in this cosmological
branch. This correspondence will prove useful in the
following sections, where we investigate the reconstruction
of the running vacuum energy density within torsion
and nonmetricity gravity.

\subsection{Nonminimal torsion-matter and nonmetricity-matter couplings}
\label{subsec:nonminimal_couplings}

An interesting extension of the above theories arises
when the geometrical sector is nonminimally coupled
to the matter Lagrangian. In particular, the
nonminimal torsion-matter coupling theory was
introduced in \cite{Harko:2014sja},
through the action
\begin{equation}
S_T=\int d^4x\,e
\left\{
\frac{-T+f_1(T)}{16\pi G}
+\left[1+\lambda f_2(T)\right]\mathcal{L}_m
\right\},
\label{eq:nonminimal_T_action}
\end{equation}
where $f_1(T)$ and $f_2(T)$ are arbitrary functions,
$\lambda$ is the coupling parameter, and
$\mathcal{L}_m$ is the matter Lagrangian.
Similarly, nonminimal coupling between nonmetricity
and matter was introduced in
\cite{Harko:2018gxr}. Adopting the same
parametrization as in the torsional case,
the corresponding action can be written as
\begin{equation}
S_Q=\int d^4x\,\sqrt{-g}
\left\{
\frac{-Q+f_1(Q)}{16\pi G}
+\left[1+\lambda f_2(Q)\right]\mathcal{L}_m
\right\}.
\label{eq:nonminimal_Q_action}
\end{equation}

Interestingly, for the spatially flat FLRW geometry,
the standard symmetric teleparallel connection,
and the perfect-fluid prescription
$\mathcal{L}_m=-\rho_m$, the two theories lead
to the same form of the background equations.
Thus, denoting by $X$ either $T$ or $Q$,
the modified Friedmann equations can be
written in the unified form
\begin{align}
3H^2 &=
8\pi G\,\mathcal{A}_X\rho_m
-\frac{1}{2}(f_1-2Xf_{1,X}),
\label{eq:nonminimal_friedmann1}\\
\dot H &=
-\frac{4\pi G(\rho_m+p_m)\mathcal{A}_X}
{\mathcal{D}_X},
\label{eq:nonminimal_friedmann2}
\end{align}
where
$
\mathcal{A}_X
=1+\lambda(f_2-2Xf_{2,X}),
\label{eq:nonminimal_A}
$
and
$
\mathcal{D}_X
=1-f_{1,X}-2Xf_{1,XX}
+16\pi G\lambda\rho_m
(f_{2,X}+2Xf_{2,XX})$,
and where subscripts denote derivatives with respect
to the corresponding geometrical scalar.

Rewriting the Friedmann equations in their
standard form, we can identify an effective
dark energy sector of geometrical origin.
In particular, its energy density and pressure
are given by
\begin{align}
\rho_{\rm DE}^{(X)}
&=
-\frac{1}{16\pi G}(f_1-2Xf_{1,X})
+\lambda\rho_m(f_2-2Xf_{2,X}),
\label{eq:nonminimal_rhoDE}\\
p_{\rm DE}^{(X)}
&=
(\rho_m+p_m)
\left(\frac{\mathcal{A}_X}{\mathcal{D}_X}-1\right)
-\rho_{\rm DE}^{(X)}.
\label{eq:nonminimal_pDE}
\end{align}
Hence, the corresponding dark energy
equation-of-state parameter becomes
\begin{equation}
w_{\rm DE}^{(X)}
=-1+
\frac{\rho_m+p_m}{\rho_{\rm DE}^{(X)}}
\left(\frac{\mathcal{A}_X}{\mathcal{D}_X}-1\right).
\label{eq:nonminimal_wDE}
\end{equation}

The nonminimal coupling between geometry and matter
can generally lead to nonconservation of the matter
energy-momentum tensor and to an effective energy
exchange between the two sectors. Nevertheless,
this behavior depends on the matter Lagrangian
and the cosmological configuration. In particular,
for the above flat FLRW equations with
$\mathcal{L}_m=-\rho_m$, one recovers the standard
matter conservation equation
$
\dot\rho_m+3H(\rho_m+p_m)=0$.
Indeed, in the considered homogeneous configurations,
the energy exchange induced by the nonminimal coupling
contains a contribution proportional to
$(\rho_m+\mathcal{L}_m)
\,d\ln[1+\lambda f_2(X)]/dt$,
which vanishes for $\mathcal{L}_m=-\rho_m$.
Thus, matter conservation follows from the adopted
fluid prescription and is not a general property
of nonminimal geometry-matter coupling theories.

These theories   offer an interesting
framework for reconstructing an effective dark
energy density with the characteristic running
vacuum dependence. However, obtaining such a
density does not necessarily imply the strict
vacuum equation of state $w_{\rm DE}=-1$,
nor the full running vacuum dynamics.
We will investigate these issues in
Sec.~\ref{sec:reconstruction}.

\subsection{Trace-coupled $f(T,\mathcal{T})$ and
$f(Q,\mathcal{T})$ gravity}
\label{subsec:trace_couplings}

Another interesting class of gravitational modifications
arises by introducing a coupling between the geometrical
sector and the trace of the matter energy-momentum tensor,
$\mathcal{T}\equiv g^{\mu\nu}T_{\mu\nu}$.
In particular, $f(T,\mathcal{T})$ gravity was introduced
in \cite{Harko:2014aja}, through the action
\begin{equation}
S_T=\frac{1}{16\pi G}\int d^4x\,e
\left[-T+f(T,\mathcal{T})\right]+S_m,
\label{eq:fTT_action}
\end{equation}
where $f(T,\mathcal{T})$ is an arbitrary function
of the torsion scalar and the matter trace.
Similarly, the corresponding extension of symmetric
teleparallel gravity was introduced in
\cite{Xu:2019sbp}. Adopting the same parametrization
as in the torsional case, its action can be written as
\begin{equation}
S_Q=\frac{1}{16\pi G}\int d^4x\,\sqrt{-g}
\left[-Q+f(Q,\mathcal{T})\right]+S_m.
\label{eq:fQT_action}
\end{equation}
For the metric signature
$(-,+,+,+)$ adopted throughout the manuscript,
the matter trace is
$\mathcal{T}=-\rho_m+3p_m$.
Furthermore, as we mentioned above, we restrict the nonmetricity formulation
to Connection I, for which $Q=6H^2$.

We mention here that the original formulations
of $f(T,\mathcal{T})$ and $f(Q,\mathcal{T})$ gravity
employ different conventions for the geometrical
scalars and the matter trace. In particular,
the original torsional formulation uses the
mostly-negative metric signature, for which
$T=-6H^2$ and $\mathcal{T}=\rho_m-3p_m$,
whereas the nonmetricity formulation uses
$Q=6H^2$ and $\mathcal{T}=-\rho_m+3p_m$.
In the common conventions adopted here,
both geometrical scalars satisfy $T=Q=6H^2$.

Furthermore, the trace-coupled field equations
depend on the adopted perfect-fluid matter
variation prescription. In particular,
$\delta\mathcal{T}/\delta g^{\mu\nu}
=T_{\mu\nu}+\Theta_{\mu\nu}$, where
$\Theta_{\mu\nu}
=g^{\alpha\beta}\delta T_{\alpha\beta}/
\delta g^{\mu\nu}$.
We adopt the perfect-fluid prescription
$\Theta_{\mu\nu}=-2T_{\mu\nu}+p_mg_{\mu\nu}$,
associated with $\mathcal{L}_m=p_m$.
With this prescription, the trace-coupling
contributions enter the homogeneous Friedmann
equations with the same sign in both theories.
In the nonmetricity case, we additionally adopt
Connection I and assume vanishing hypermomentum.

Introducing the notation $X=T$ or $Q$, with
$X=6H^2$, the modified Friedmann equations can
be written as
\begin{align}
H^2
&=
\frac{8\pi G}{3}\rho_m
-\frac{1}{6}(f-2Xf_X)
+\frac{1}{3}f_{\mathcal{T}}
(\rho_m+p_m),
\label{eq:trace_friedmann1}\\
\dot H
&=
-\frac{4\pi G(\rho_m+p_m)}
{\mathcal{D}_X}
\left(
1+\frac{  f_{\mathcal{T}}}{8\pi G}
\right),
\label{eq:trace_friedmann2}
\end{align}
where
$
\mathcal{D}_X
=
1-f_X-2Xf_{XX}
+H\frac{d\rho_m}{dH}
(1-3c_s^2)f_{X\mathcal{T}}$.
Here $c_s^2=dp_m/d\rho_m$ denotes the
adiabatic sound speed of the barotropic matter
fluid, while subscripts indicate partial
derivatives with respect to the corresponding
arguments. The derivative $d\rho_m/dH$ is
evaluated along the cosmological solution.

Rewriting the above equations in their standard
form, we can identify an effective dark energy
sector of gravitational origin. In particular,
its energy density and pressure are given by
\begin{align}
\rho_{\rm DE}^{(X)}
&=
\frac{1}{16\pi G}
\left[
-f+2Xf_X
+2  f_{\mathcal{T}}(\rho_m+p_m)
\right],
\label{eq:trace_rhoDE}\\
p_{\rm DE}^{(X)}
&=
(\rho_m+p_m)
\left[
\frac{1+  f_{\mathcal{T}}/8\pi G}
{\mathcal{D}_X}-1
\right]
-\rho_{\rm DE}^{(X)}.
\label{eq:trace_pDE}
\end{align}
Hence, the corresponding dark energy
equation-of-state parameter becomes
\begin{equation}
w_{\rm DE}^{(X)}
=
-1+
\frac{\rho_m+p_m}{\rho_{\rm DE}^{(X)}}
\left[
\frac{1+  f_{\mathcal{T}}/8\pi G}
{\mathcal{D}_X}-1
\right].
\label{eq:trace_wDE}
\end{equation}

An important feature of these theories is that
the explicit coupling between geometry and the
matter trace generally leads to nonconservation
of the matter energy-momentum tensor. Consequently,
the matter and effective dark energy sectors
can exchange energy during the cosmological
evolution. The precise form of this exchange
depends on the gravitational Lagrangian and
the matter prescription.

In particular, for pressureless matter,
$p_m=0$, the strict vacuum equation of state
$p_{\rm DE}^{(X)}=-\rho_{\rm DE}^{(X)}$
requires
\begin{equation}
\mathcal{D}_X
=
1+\frac{  f_{\mathcal{T}}}{8\pi G},
\label{eq:trace_vacuum_condition}
\end{equation}
on a regular branch with nonvanishing matter
density. Under this condition,
Eq.~(\ref{eq:trace_friedmann2}) reduces to
\begin{equation}
\dot H=-4\pi G\rho_m.
\label{eq:trace_vacuum_hdot}
\end{equation}
Thus, unlike the minimally coupled theories,
the trace-coupled formulations possess the
necessary freedom to reproduce the acceleration
equation of running vacuum cosmology while
maintaining a time-dependent effective vacuum
density.

Indeed, as discussed in Sec.~\ref{sec:rvm},
a time-dependent vacuum energy density with
$w_{\rm DE}=-1$ requires a nonvanishing
interaction with matter for constant $G$. Thus, trace-coupled gravity provides a
framework
in which such an interaction can emerge from
the gravitational field equations. In Sec. \ref{subsec:rvm_trace_quartic}
we will investigate whether suitable gravitational
Lagrangians can reproduce simultaneously the
running vacuum density, its strict equation
of state, and the corresponding homogeneous
cosmological evolution.

\section{Running vacuum density reconstruction in torsional and nonmetricity gravity}
\label{sec:reconstruction}

In this section, we investigate whether torsional and
nonmetricity gravity can reproduce the characteristic
running vacuum density, including the constant, $H^2$,
and $H^4$ contributions. We first perform the
reconstruction in minimally coupled $f(T)$ and $f(Q)$
theories and examine the resulting cosmological
evolution and observational constraints. We then
extend the analysis to nonminimal geometry-matter
couplings. Throughout this section, we focus on
reproducing the effective vacuum density, without
imposing the strict equation of state
$w_{\rm DE}=-1$.

\subsection{Reconstruction in $f(T)$ and $f(Q)$ gravity}
\label{subsec:rvm_fT_fQ}
\subsubsection{Analytical reconstruction}
\label{subsubsec:rvm_analytical}

We   seek minimally coupled gravitational models whose effective dark energy
density reproduces the running vacuum form of Eq.~(\ref{eq:rvm_density}). In $f(T)$
gravity   the effective density is given by (\ref{eq:fT_rhoDE}). Hence, equating
the two expressions, and using $T=6H^2$,  gives $2Tf_T-f=6c_0+\nu T+(\gamma/6)T^2$,
whose general solution is $f(T)=-6c_0+\nu T+(\gamma/18)T^2+C_T\sqrt{T}$. Since the
last term does not affect the homogeneous field equations, we set $C_T=0$ and obtain
\begin{equation}
 f(T)=-6c_0+\nu T+\frac{\gamma}{18}T^2.
\label{eq:rvm_fT_polynomial}
\end{equation}

The exact dark energy density equality does not imply running vacuum behavior. In
particular, from Eq.~(\ref{eq:fT_pDE}), the reconstructed model yields
\begin{equation}
 w_{\rm DE}=
 \frac{-c_0+\gamma H^4}
 {(1-\nu-2\gamma H^2)(c_0+\nu H^2+\gamma H^4)},
\label{eq:rvm_fT_eos}
\end{equation}
(we focus on the
cases where the denominator is nonzero)
which generally differs from $-1$, unlike genuine running vacuum.

Moreover, as we mentioned in subsection   \ref{subsec:fT_fQ}, in $f(T)$ and $f(Q)$
gravity
matter  is separately conserved, which leads to $\rho_m^{f(T)}=\rho_{m0}(1+z)^3$,
whereas running vacuum cosmology requires the matter-vacuum exchange
$\dot\rho_m+3H\rho_m=3H(\nu+2\gamma H^2)\rho_m$ derived in Sec.~\ref{sec:rvm}.
Hence, although both scenarios obey the same algebraic relation
$\rho_m(H)=3[(1-\nu)H^2-c_0-\gamma H^4]/(8\pi G)$, their different conservation laws
produce distinct expansion histories. Indeed, for  $f(T)$ gravity we have
\begin{equation}
 \dot H_{f(T)}=-\frac{3}{2}\,
 \frac{(1-\nu)H^2-c_0-\gamma H^4}
 {1-\nu-2\gamma H^2},
\label{eq:rvm_fT_hdot}
\end{equation}
whereas in running vacuum model (RVM) we acquire $\dot H_{\rm
RVM}=-\frac{3}{2}[(1-\nu)H^2-c_0-\gamma H^4]$, without the additional denominator.
Thus, matching the density as a function of $H$ is insufficient to match its
evolution in time or redshift.

To express this difference explicitly, we introduce $E(z)=H(z)/H_0$, $C=c_0/H_0^2$,
and $\Gamma=\gamma H_0^2$, with $C=1-\Omega_{m0}-\nu-\Gamma$, and where a subscript
``0'' marks the values of a quantity at present. The  $f(T)$ gravity
model obeys
\begin{equation}
 (1-\nu)E_{f(T)}^2-C-\Gamma E_{f(T)}^4
 =\Omega_{m0}(1+z)^3,
\label{eq:rvm_fT_Ez}
\end{equation}
whereas the running vacuum evolution is governed by
\begin{equation}
 \frac{dE_{\rm RVM}^2}{dz}=
 \frac{3}{1+z}\left[(1-\nu)E_{\rm RVM}^2-C
 -\Gamma E_{\rm RVM}^4\right],
\label{eq:rvm_comparison_Ez}
\end{equation}
with $E(0)=1$ in both cases. The regular $f(T)$ branch continuously connected to the
standard solution is
\begin{equation}
 E_{f(T)}^2(z)=
 \frac{2[C+\Omega_{m0}(1+z)^3]}
 {1-\nu+\sqrt{(1-\nu)^2-4\Gamma[C+\Omega_{m0}(1+z)^3]}},
\label{eq:rvm_fT_Ez_solution}
\end{equation}
valid for either sign of $\Gamma$ on the branch normalized by $E(0)=1$, provided
$1-\nu-2\Gamma>0$ and the background remains real  and regular over the range
considered.

An important consequence is that the $f(T)$ expansion depends on only two
combinations of the three parameters $\Omega_{m0}$, $\nu$, and $\Gamma$:
\begin{align}
 E_{f(T)}^2-1&=\alpha[(1+z)^3-1]
 +\beta(E_{f(T)}^4-1),
\label{eq:rvm_fT_Ez_reduced}\\
 \alpha&=\frac{\Omega_{m0}}{1-\nu},\qquad
 \beta=\frac{\Gamma}{1-\nu}.
\label{eq:rvm_fT_effective_parameters}
\end{align}
Background data can therefore constrain $\alpha$ and $\beta$, but cannot
independently determine $\Omega_{m0}$, $\nu$, and $\Gamma$ in this model. No
analogous exact degeneracy occurs for the RVM evolution equation. In both cases,
$\Omega_{\rm DE}(z)=\nu+C/E^2(z)+\Gamma E^2(z)$, but the distinct functions $E(z)$
imply different redshift evolutions.

Finally, we mention that all the above discussion is valid
for $f(Q)$ gravity with Connection I, under     $T\rightarrow Q$. The
corresponding reconstruction is thus
\begin{equation}
 \tilde f(Q)=-6c_0+\nu Q+\frac{\gamma}{18}Q^2,
\label{eq:rvm_fQ_solution}
\end{equation}
with identical homogeneous evolution to the torsional model. Hence, both minimally
coupled theories reproduce the running vacuum density, but not, in general,
its strict equation of state or interacting dynamics.

We close this analysis by mentioning that
reconstructing a more general running vacuum
energy density in $f(T)$ or $f(Q)$ gravity is
straightforward, since $H^2=T/6=Q/6$. For instance, terms of the form $f(T)=a_nT^n$
(and similarly for
 $T\rightarrow Q$) produce terms $ \propto H^{2n}$, when $n\neq1/2$. More generally,
one can reproduce a general $\rho_{\rm DE}(H)$
 using
\begin{equation}
 f(T)=\sqrt{T}\left[C_T+8\pi G\int^T
 \frac{\rho_{\rm DE}(u)}{u^{3/2}}\,du\right],
\label{eq:rvm_fT_general_reconstruction}
\end{equation}
(and similarly for
 $T\rightarrow Q$). However, as we mentioned, these models will not in general  have
$ w_{\rm DE}=-1$.

\begin{figure}[ht!]
\includegraphics[width=0.42\textwidth]{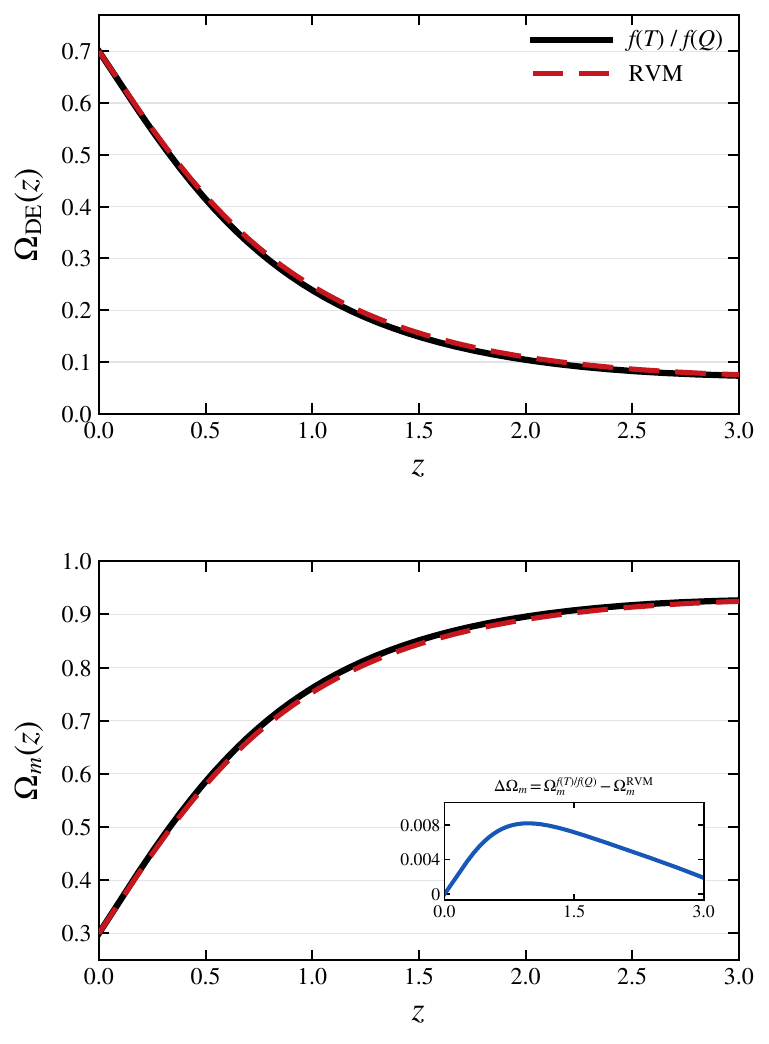}
\caption{{\it{
Evolution of the effective dark energy density
parameter $\Omega_{\rm DE}(z)$ (upper panel)
and the matter density parameter $\Omega_m(z)$
(lower panel) in minimally coupled $f(T)$
and $f(Q)$ gravity (black solid lines) and
running vacuum cosmology (red dashed lines).
The inset shows the difference
$\Delta\Omega_m=
\Omega_m^{f(T)/f(Q)}-\Omega_m^{\rm RVM}$.
Although the models possess identical functional
dependence $\rho_{\rm DE}(H)$, their different
matter conservation laws lead to distinct
redshift evolutions. We use
$\Omega_{m0}=0.30$, $\nu=0.02$,
$\gamma H_0^2=10^{-3}$, and
$c_0/H_0^2=0.679$.
}}}
\label{fig:density_matching}
\end{figure}

\subsubsection{Comparison under cosmological evolution and observational constraints}
\label{subsubsec:rvm_observational}

In the previous paragraph, within $f(T)$ and $f(Q)$ gravity we
reconstructed the running vacuum energy density expression, but neither $ w_{\rm
DE}=-1$
nor interacting behavior. Thus, although the two evolutions will be   close, they
will not be identical.

We first compare the background evolution of the minimally
coupled $f(T)$ or $f(Q)$ reconstruction and running vacuum scenario, for the same
illustrative
parameters, using the redshift $z=-1+1/a$ as the independent variable.
As shown in Fig.~\ref{fig:density_matching},
the density parameters evolve similarly, with
small deviations at intermediate redshifts arising
from the different matter conservation laws.
Although the effective dark energy densities have
the same functional dependence on $H$, their
redshift evolution differs because the corresponding
expansion histories are not identical.

\begin{figure}[ht!]
\includegraphics[width=0.42\textwidth]{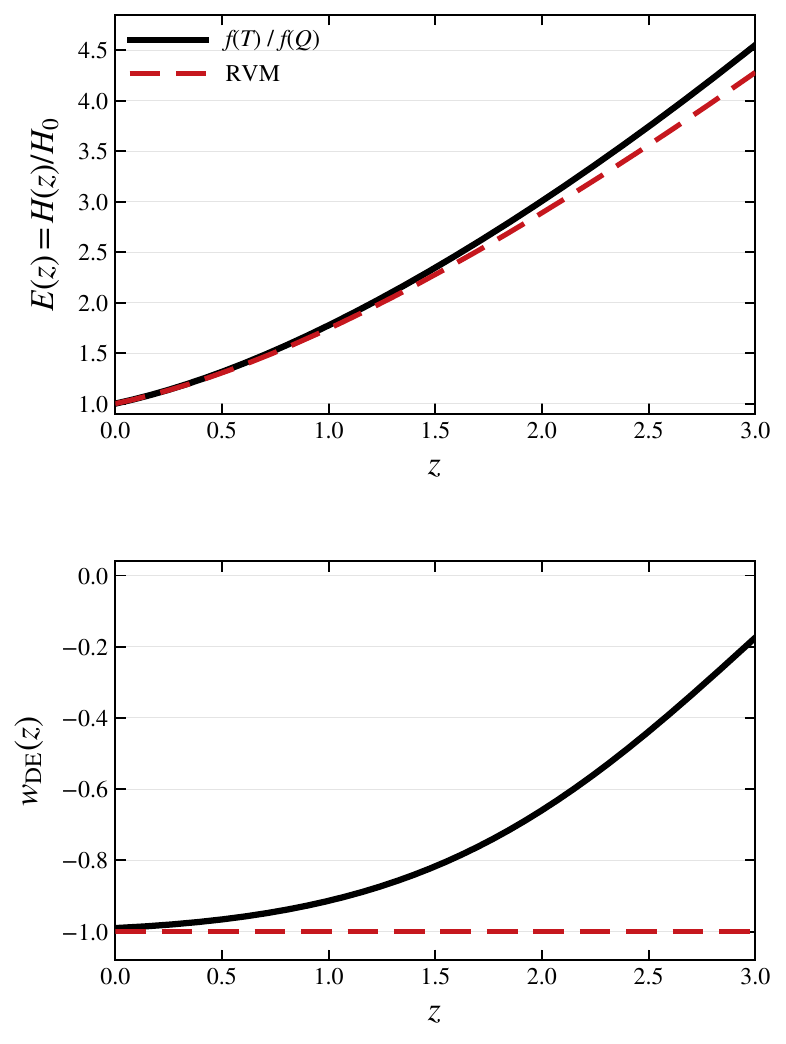}
\caption{{\it{
Comparison of the normalized Hubble expansion
rate $E(z)=H(z)/H_0$ (upper panel) and the
effective dark energy equation-of-state
parameter $w_{\rm DE}(z)$ (lower panel)
in minimally coupled $f(T)$ and $f(Q)$
gravity (black solid lines) and running
vacuum cosmology (red dashed lines).
Despite reproducing the same
$\rho_{\rm DE}(H)$, the minimally coupled
theories exhibit a different expansion
history and an evolving effective dark
energy equation of state, whereas running
vacuum cosmology satisfies $w_{\rm DE}=-1$
identically. We use $\Omega_{m0}=0.30$,
$\nu=0.02$, $\gamma H_0^2=10^{-3}$,
and $c_0/H_0^2=0.679$.
}}}
\label{fig:background_comparison}
\end{figure}

In Fig. \ref{fig:background_comparison} we depict the evolution of the expansion
rate and of  $w_{\rm DE}$. In the reconstructed models we  observe a  slightly
higher
expansion rate and an evolving
$w_{\rm DE}$, whereas running vacuum preserves $w_{\rm DE}=-1$. Note that
the displayed parameters are illustrative rather than observational
best-fit values.

To assess the observational implications,
we perform a joint fit to the observational data.  We use 1590 unbinned
Pantheon+ Type Ia supernova light curves with $z_{\rm HD}>0.01$
and their statistical and systematic covariance
\cite{Brout:2022vxf}, 31 cosmic chronometer measurements
\cite{Jimenez:2001gg,Moresco:2016mzx}, and 13 DESI DR2 BAO observables
($D_M/r_d$, $D_H/r_d$, and $D_V/r_d$), including the
Lyman-$\alpha$ measurements \cite{DESI:2025zgx}. The sound horizon $r_d$ is fitted
freely. We adopt $\mathcal{L}_{\rm tot}\propto
\exp(-\chi^2_{\rm tot}/2)$, with
\begin{equation}
\chi^2_{\rm tot}=\chi^2_{\rm SNIa}+\chi^2_{\rm CC}+\chi^2_{\rm BAO},
\label{eq:rvm_total_chi2}
\end{equation}
using the published supernova and BAO covariance matrices and the
quoted chronometer uncertainties. The supernova zero point is
profiled analytically. We employ flat priors, allowing either sign
of the running parameters, and sample the posterior using
\texttt{emcee} \cite{Foreman-Mackey:2012any} with $30000$
steps per walker. Finally, radiation is neglected in this late-time analysis.

For $f(T)$ gravity we fit $\{\alpha,\beta,H_0,r_d\}$, where
$\alpha=\Omega_{m0}/(1-\nu)$ and $\beta=\Gamma/(1-\nu)$,
with $\Gamma=\gamma H_0^2$, since the background cannot
independently constrain $\Omega_{m0}$, $\nu$, and $\Gamma$.
For RVM the sampled parameters are
$\{\Omega_{m0},\nu,\Gamma,H_0,r_d\}$.

\begin{figure*}[ht!]
\includegraphics[width=0.80\textwidth]{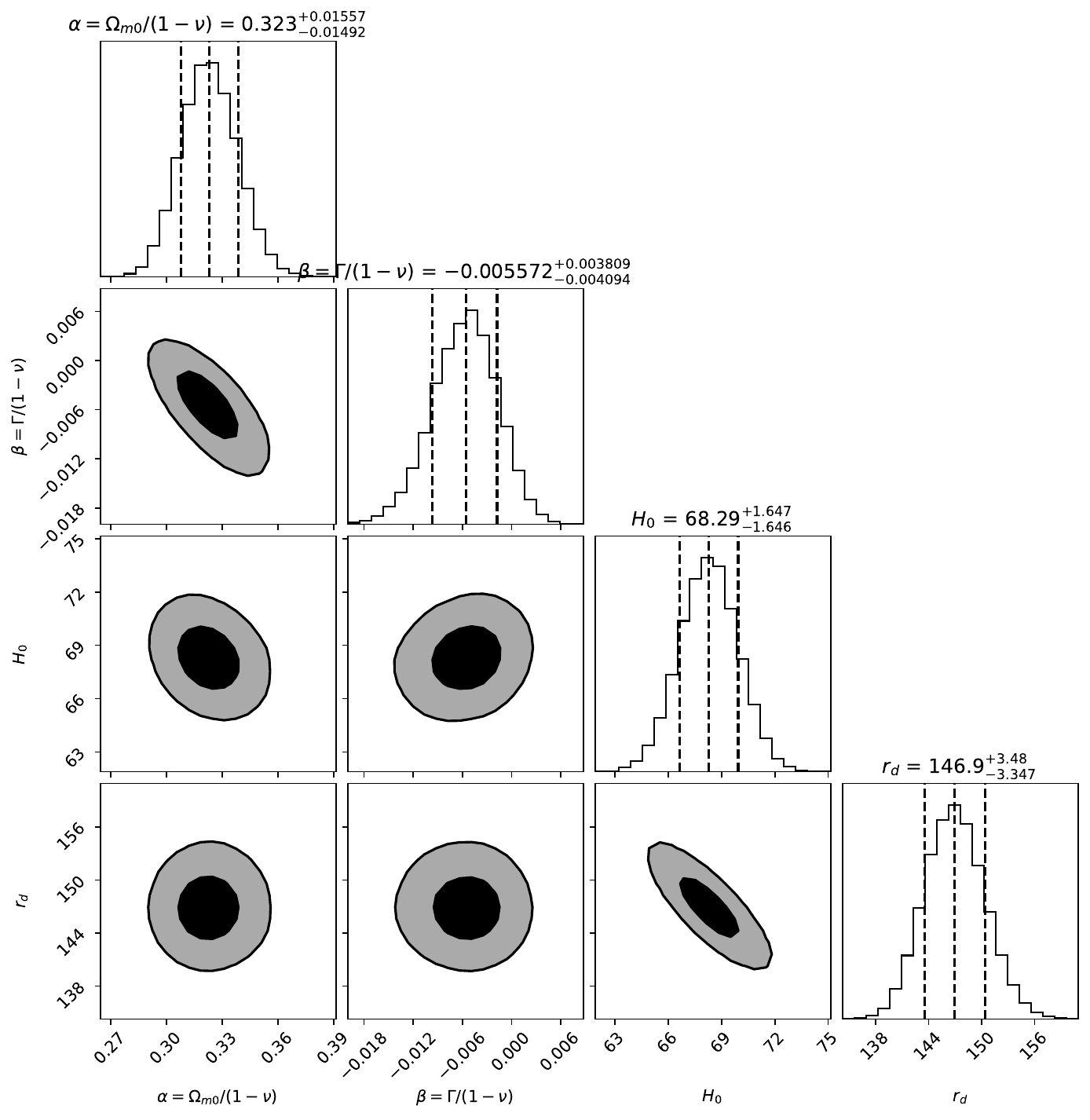}
\caption{{\it{Marginalized posterior distributions and
two-dimensional $1\sigma$ and $2\sigma$
credible contours for the minimally
coupled $f(T)$ gravitational reconstruction,
obtained from the joint Pantheon+,
cosmic chronometer, and DESI DR2 BAO
analysis. The independent background
parameters are
$\alpha=\Omega_{m0}/(1-\nu)$,
$\beta=\Gamma/(1-\nu)$,
$H_0$, and $r_d$, where
$\Gamma=\gamma H_0^2$.
The results also describe the corresponding
$f(Q)$ reconstruction with Connection I
at the homogeneous cosmological level.
}}}
\label{fig:FTcontours}
\end{figure*}

In  Fig. \ref{fig:FTcontours} we show the results for the $f(T)$ scenario. In
particular, it gives the   marginalized
constraints (68\% confidence):
$\alpha=0.3230^{+0.0156}_{-0.0149}$,
$\beta=-0.00557^{+0.00381}_{-0.00409}$,
$H_0=68.29^{+1.65}_{-1.65}\,{\rm km\,s^{-1}\,Mpc^{-1}}$,
and $r_d=146.9^{+3.48}_{-3.35}\,{\rm Mpc}$.
The preference for negative $\beta$ is mild, with $\beta=0$
still allowed; $\alpha$ and $\beta$ are anticorrelated.
We mention that all these results
 also apply to the $f(Q)$ reconstruction with
Connection I.

\begin{figure*}[ht!]
\includegraphics[width=0.80\textwidth]{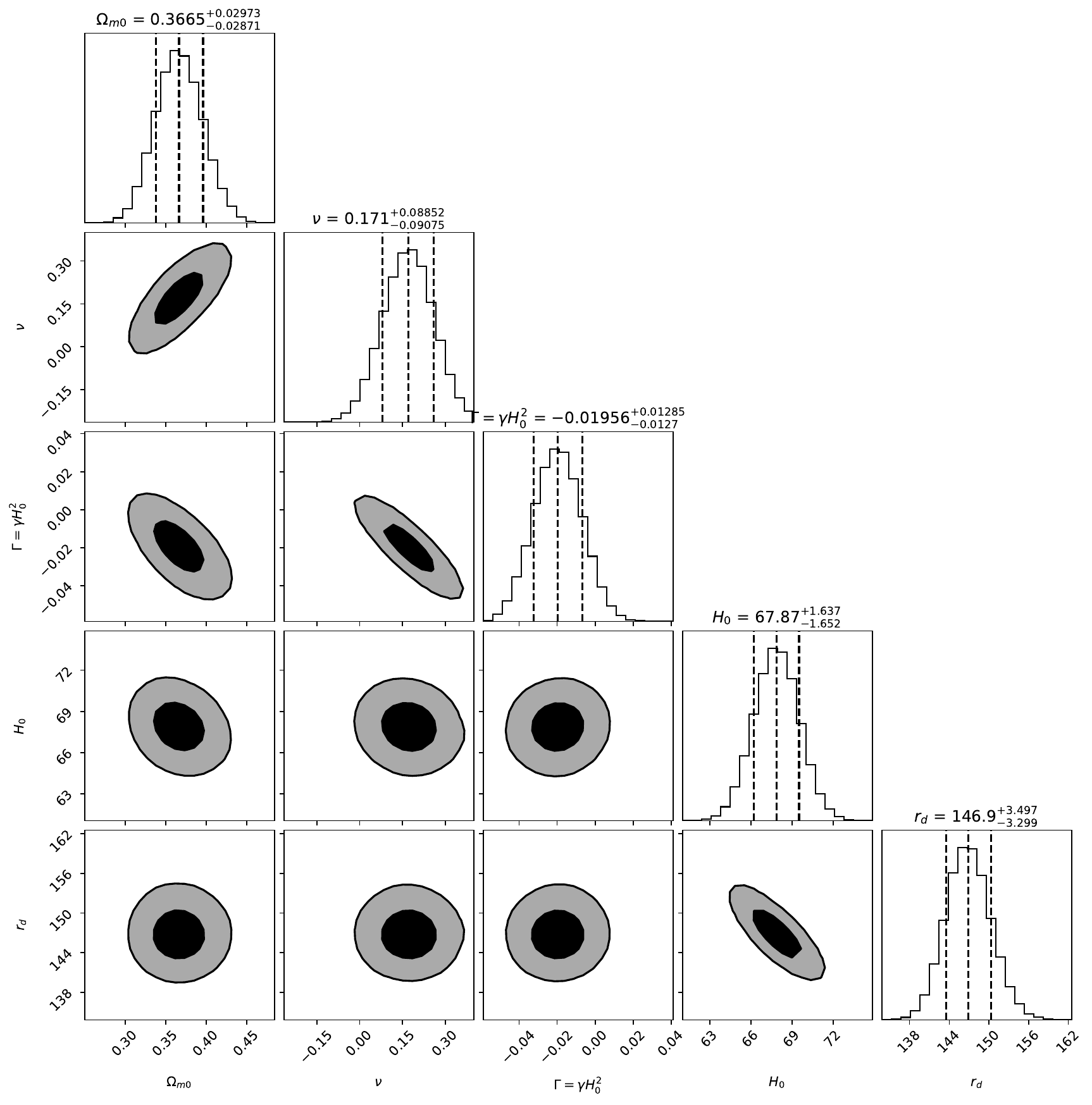}
\caption{{\it{Marginalized posterior distributions and
two-dimensional $1\sigma$ and $2\sigma$
credible contours for the running
vacuum model, obtained from the joint
Pantheon+, cosmic chronometer, and
DESI DR2 BAO analysis.
The independent parameters are
$\Omega_{m0}$, $\nu$,
$\Gamma=\gamma H_0^2$, $H_0$,
and $r_d$. The correlations between
the matter density parameter and
the running vacuum coefficients
illustrate the distinct background
dynamics of the interacting vacuum
scenario.
}}}
\label{fig:RVcontours}
\end{figure*}

In Fig.~\ref{fig:RVcontours} we present the corresponding results for the
 running vacuum scenario. As we can see, it yields (68\% confidence)
$\Omega_{m0}=0.3665^{+0.0297}_{-0.0287}$,
$\nu=0.1710^{+0.0885}_{-0.0908}$,
$\Gamma=-0.01956^{+0.01285}_{-0.01270}$,
$H_0=67.87^{+1.64}_{-1.65}\,{\rm km\,s^{-1}\,Mpc^{-1}}$,
and $r_d=146.9^{+3.50}_{-3.30}\,{\rm Mpc}$.
The running parameters are strongly correlated with $\Omega_{m0}$,
while $H_0$ and $r_d$ remain close to the $f(T)$ estimates.

For a direct model comparison we apply the
Akaike and Bayesian information criteria
\cite{Liddle:2007fy}, namely
\begin{equation}
{\rm AIC}=\chi^2_{\min}+2k,\qquad
{\rm BIC}=\chi^2_{\min}+k\ln N,
\label{eq:rvm_aic_bic}
\end{equation}
where $N=1590+31+13=1634$ and $k=5$ for $f(T)$/$f(Q)$
or $k=6$ for RVM, including the profiled supernova zero point. The BIC uses the
nominal number of correlated data entries.
The results are given in Table~\ref{tab:rvm_model_comparison}.
\begin{table}[ht!]
\centering
\caption{Statistical comparison between the minimally
coupled $f(T)/f(Q)$ reconstruction and running vacuum
cosmology using the joint Pantheon+ + CC + DESI DR2
BAO dataset. The parameter count $k$ includes the
profiled supernova zero point, and $N=1634$.
The differences are defined as the $f(T)/f(Q)$
value minus the RVM value. The information criteria
are quoted up to the common likelihood normalization.}
\label{tab:rvm_model_comparison}
\begin{tabular}{lcccc}
\hline\hline
Model & $k$ & $\chi^2_{\min}$ & AIC & BIC \\
\hline
$f(T)$/$f(Q)$ & 5 & 1428.824 & 1438.824 & 1465.818 \\
RVM          & 6 & 1424.975 & 1436.975 & 1469.368 \\
\hline
Difference   & - & $+3.849$ & $+1.849$ & $-3.550$ \\
\hline\hline
\end{tabular}
\end{table}

As we can see, RVM improves the minimum fit by $\Delta\chi^2_{\min}=3.849$
(relative to $f(T)$/$f(Q)$), but has one additional parameter.
Accordingly, $\Delta{\rm AIC}=+1.849$ weakly favors RVM,
whereas $\Delta{\rm BIC}=-3.550$ favors the geometrical
reconstruction (note that the differences are defined as $f(T)$/$f(Q)$ minus
RVM). Neither criterion provides a decisive preference.

Finally, we mention here that our purpose
was to show that even with the same dark energy density form,
the $f(T)$/$f(Q)$ reconstructions and RVM
exhibit slightly different cosmological evolution,
leading to distinguishable observational predictions. Our purpose was not to compare
with the $\Lambda$CDM paradigm, a comparison that has been separately performed in
detail in the literature.

\subsection{Reconstruction in nonminimally matter-coupled theories}
\label{subsec:rvm_nonminimal}

We now proceed to the consideration of gravitational theories with nonminimal
geometry-matter couplings, described by the functions
$f_1(T)$ and $f_2(T)$ introduced in
Sec.~\ref{subsec:nonminimal_couplings}.
Since these theories contain minimally coupled $f(T)$
gravity as a particular limit, the running vacuum density
$\rho_{\rm DE}(H)\propto c_0+\nu H^2+\gamma H^4$
can be reconstructed without difficulty. The relevant
question is whether the additional matter coupling
can also reproduce the strict vacuum equation of state
$p_{\rm DE}=-\rho_{\rm DE}$ while maintaining a
time-dependent effective energy density.

From the modified Friedmann equations
(\ref{eq:nonminimal_friedmann1}),(\ref{eq:nonminimal_friedmann2}),
of
Sec.~\ref{subsec:nonminimal_couplings}, we have
\begin{equation}
\rho_{\rm DE}+p_{\rm DE}
=
(\rho_m+p_m)
\left(\frac{\mathcal{A}_T}{\mathcal{D}_T}-1\right).
\label{eq:rvm_nonminimal_rhop}
\end{equation}
Hence, for a regular cosmological
evolution with $\rho_m+p_m\neq0$, the vacuum
condition requires
\begin{equation}
\mathcal{A}_T=\mathcal{D}_T.
\label{eq:rvm_nonminimal_condition}
\end{equation}
Although this relation constrains the functions
$f_1(T)$ and $f_2(T)$, a more fundamental restriction
arises from the conservation equations.

Indeed, for the matter prescription
$\mathcal{L}_m=-\rho_m$ adopted here, the
homogeneous matter sector is separately conserved,
namely $
\dot\rho_m+3H(\rho_m+p_m)=0$.
Since the effective dark energy density and pressure
are defined by rewriting the modified Friedmann
equations in their standard form, total conservation
then implies
$
\dot\rho_{\rm DE}
+3H(\rho_{\rm DE}+p_{\rm DE})=0$.
Consequently, imposing $p_{\rm DE}=-\rho_{\rm DE}$
necessarily gives $\dot\rho_{\rm DE}=0$, independently
of the particular forms of $f_1(T)$ and $f_2(T)$.

However, for the running vacuum density of
Eq.~(\ref{eq:rvm_density}), the  condition  $\dot\rho_{\rm DE}=0$  gives
\begin{equation}
H\dot H(\nu+2\gamma H^2)=0,
\label{eq:rvm_nonminimal_running_constraint}
\end{equation}
and
therefore  for a nontrivial cosmological evolution with a
continuously varying Hubble parameter, it follows that
$\nu=\gamma=0$. Hence, the effective dark energy
reduces to a cosmological constant,
$\rho_{\rm DE}=3c_0/(8\pi G)$, and the resulting
background evolution coincides with the $\Lambda$CDM paradigm.

In summary, although the considered nonminimal
torsion-matter coupling allows the reconstruction
of the running vacuum density, it cannot reproduce
simultaneously its time dependence and the strict
vacuum equation of state. The same conclusion holds
for the corresponding nonminimal $f(Q)$ theory with
Connection I, whose homogeneous equations have
the same structure, and can be obtained from the above expressions under
$T\rightarrow Q$. However, we mention that this result applies to the
adopted matter prescription and cosmological
configuration, rather than constituting a general
no-go theorem for arbitrary geometry-matter
couplings. It also highlights the importance of
matter nonconservation for a genuine running
vacuum realization, motivating the trace-coupled
theories examined in the following section.

\section{Exact running vacuum reconstruction in trace-coupled gravity}
\label{sec:exact_rvm}

In this section, we investigate whether trace-coupled
$f(T,\mathcal{T})$ and $f(Q,\mathcal{T})$ gravity can
reproduce the complete running vacuum dynamics,
beyond the effective density matching of the
previous section. We first derive the conditions
required to recover simultaneously the running
vacuum density, the strict equation of state
$w_{\rm DE}=-1$, and the characteristic matter-vacuum
interaction. Then we  obtain explicit linear
reconstructions for the constant and $H^2$ terms,
and extend the analysis to the nonlinear $H^4$
contribution, examining the existence and regularity
of the resulting cosmological solutions.

\subsection{Conditions for exact running vacuum equivalence}
\label{subsec:trace_exact_conditions}

We now examine whether trace-coupled $f(T,\mathcal{T})$
and $f(Q,\mathcal{T})$ gravity can reproduce the complete
running vacuum dynamics. This requires simultaneously
\begin{equation}
\rho_{\rm DE}(H)=
\frac{3}{8\pi G}
\left(c_0+\nu H^2+\gamma H^4\right),
\qquad
p_{\rm DE}=-\rho_{\rm DE}.
\label{eq:trace_recon_requirements}
\end{equation}
We consider pressureless matter, for which $\mathcal{T}=-\rho_m$,
and introduce $X=T$ or $Q$, with $X=6H^2$ and
$K=(8\pi G)^{-1}$. In the conventions of
Sec.~\ref{subsec:trace_couplings}, the modified
Friedmann equations take the unified form
\begin{align}
H^2
&=
\frac{\rho_m}{3K}
-\frac{1}{6}(f-2Xf_X)
+\frac{1}{3}f_{\mathcal{T}}\rho_m,
\label{eq:trace_recon_friedmann1}\\
\dot H
&=
-\frac{\rho_m}{2K}
\frac{1+ Kf_{\mathcal{T}}}
{\mathcal{D}_X},
\label{eq:trace_recon_friedmann2}
\end{align}
where
\begin{equation}
\mathcal{D}_X
=
1-f_X-2Xf_{XX}
+2X\frac{d\rho_m}{dX}f_{X\mathcal{T}}.
\label{eq:trace_recon_D}
\end{equation}
The derivative $d\rho_m/dX$ is evaluated along the
cosmological solution, and Connection I is adopted
in the nonmetricity case.

Rewriting the field equations in their standard
form gives
$\rho_{\rm DE}+p_{\rm DE}
=\rho_m[(1+  Kf_{\mathcal{T}})/
\mathcal{D}_X-1]$.
Hence, for nonvanishing $\rho_m$, the strict vacuum
equation of state requires
\begin{equation}
\mathcal{D}_X=1+ Kf_{\mathcal{T}},
\label{eq:trace_recon_vacuum_condition}
\end{equation}
on regular branches where both sides are nonzero.
Under this condition, Eq.~(\ref{eq:trace_recon_friedmann2})
reduces to $\dot H=-4\pi G\rho_m$, exactly as in the running vacuum scenario.

Furthermore, imposing the density of
Eq.~(\ref{eq:trace_recon_requirements}) in the first
Friedmann equation determines the required matter
density as
\begin{equation}
\rho_m(X)
=
K\left[
\frac{1-\nu}{2}X
-3c_0-\frac{\gamma}{12}X^2
\right].
\label{eq:trace_recon_matter}
\end{equation}
Combining this expression with
$\dot H=-4\pi G\rho_m$, we recover
$\dot H=-\frac{3}{2}[(1-\nu)H^2-c_0-\gamma H^4]$,
together with the characteristic interaction
$\dot\rho_m+3H\rho_m
=3H(\nu+2\gamma H^2)\rho_m$.
Thus, both the expansion history and the
matter-vacuum energy exchange coincide with those of the RVM scenario,
without introducing an additional interaction term.

The exact reconstruction therefore reduces to
determining $f(X,\mathcal{T})$ satisfying
Eqs.~(\ref{eq:trace_recon_friedmann1}) and
(\ref{eq:trace_recon_vacuum_condition}), with
$\rho_m(X)$ given by Eq.~(\ref{eq:trace_recon_matter}). In the following, we first
obtain explicit linear solutions for the
constant and $H^2$ contributions, and subsequently
extend the reconstruction to the nonlinear $H^4$ term.

\subsection{Exact reconstruction of the $\rho_{\rm DE}(H)=
\frac{3}{8\pi G}(c_0+\nu H^2)$ running vacuum model}
\label{subsec:rvm_trace_linear}

We first consider the running vacuum scenario with
$\gamma=0$, namely
\begin{equation}
\rho_{\rm DE}(H)=
\frac{3}{8\pi G}(c_0+\nu H^2).
\label{eq:trace_linear_target}
\end{equation}
In this case, the matter density required by
Eq.~(\ref{eq:trace_recon_matter}) reduces to
$\rho_m(X)=K[(1-\nu)X/2-3c_0]$.
We will show that this cosmological evolution
can be reproduced exactly through linear
trace-coupled gravitational Lagrangians.

We start with $f(T,\mathcal{T})$ gravity and
consider the ansatz
\begin{equation}
f(T,\mathcal{T})
=
aT-8\pi G a\mathcal{T}+b,
\label{eq:trace_linear_fT}
\end{equation}
where $a$ and $b$ are constants. Substituting
this expression into the first modified Friedmann
equation (\ref{eq:trace_recon_friedmann1}),
with $\mathcal{T}=-\rho_m$, gives
\begin{equation}
\rho_m=
\frac{K}{2-3a}\left[(1-a)T+b\right],
\label{eq:trace_linear_rhom}
\end{equation}
and the corresponding effective dark energy density
is then
\begin{equation}
\rho_{\rm DE}=
\frac{3}{8\pi G}
\left[
-\frac{b}{3(2-3a)}
-\frac{a}{2-3a}H^2
\right].
\label{eq:trace_linear_density}
\end{equation}
Comparing with Eq.~(\ref{eq:trace_linear_target}), we identify
$\nu=-a/(2-3a)$ and
$c_0=-b/[3(2-3a)]$.
Thus, $a=-2\nu/(1-3\nu)$ and
$b=-6c_0/(1-3\nu)$.
Therefore, the gravitational Lagrangian can be
expressed directly in terms of the running
vacuum parameters as
\begin{equation}
f(T,\mathcal{T})=
-\frac{2\nu}{1-3\nu}T
+\frac{16\pi G\nu}{1-3\nu}\mathcal{T}
-\frac{6c_0}{1-3\nu}.
\label{eq:trace_linear_fT_final}
\end{equation}
Hence, the desired constant and $H^2$
contributions are reproduced exactly.

However, the crucial feature of this reconstruction
is that it also satisfies the strict vacuum
equation of state. Indeed, for the above
Lagrangian, Eq.~(\ref{eq:trace_recon_D}) yields
\begin{equation}
\mathcal{D}_T=1-a=1+Kf_{\mathcal{T}}=\frac{1-\nu}{1-3\nu},
\label{eq:trace_linear_DT}
\end{equation}
which coincides with the vacuum condition
(\ref{eq:trace_recon_vacuum_condition}). Consequently, the second
modified Friedmann equation becomes
$\dot H=-4\pi G\rho_m$, implying
$p_{\rm DE}=-\rho_{\rm DE}$ and
$w_{\rm DE}=-1$ throughout the cosmological
evolution. Hence, the reconstructed theory
reproduces not only the running vacuum density
but also its characteristic vacuum behavior.

\begin{figure}[ht!]
\includegraphics[width=0.42\textwidth]{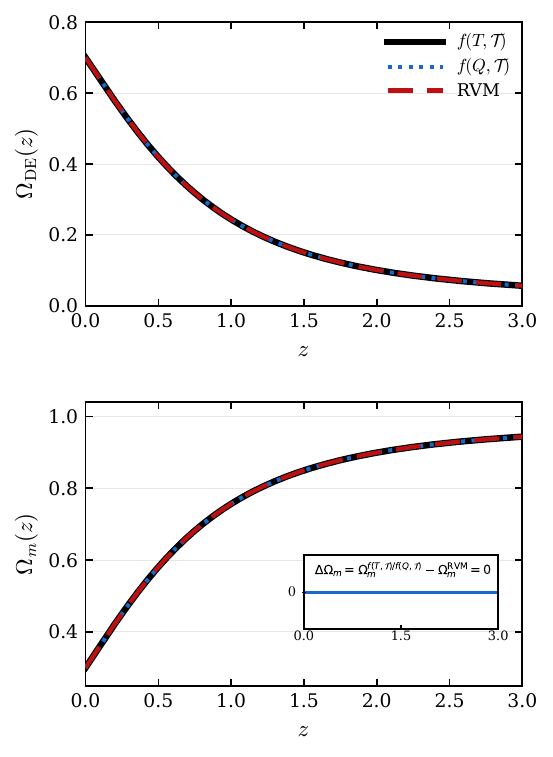}
\caption{{\it{
Evolution of the effective dark energy density
parameter $\Omega_{\rm DE}(z)$ (upper panel)
and the matter density parameter $\Omega_m(z)$
(lower panel) for the reconstructed
$f(T,\mathcal{T})$ gravity (black solid line),
$f(Q,\mathcal{T})$ gravity with Connection I
(blue dotted line), and running vacuum cosmology
(red dashed line), for $\gamma=0$.
The three models have identical parameters
and initial conditions, and their curves
coincide exactly. The inset shows the vanishing
difference $\Delta\Omega_m=0$, confirming the
exact reconstruction of the matter evolution. We use $\Omega_{m0}=0.30$, $\nu=0.02$,
$\gamma=0$, and $c_0/H_0^2=0.68$.
}}}
\label{quadraticdensity}
\end{figure}

Furthermore, the time dependence of the effective
vacuum density implies a corresponding exchange
of energy with matter. In particular,
using (\ref{eq:trace_linear_rhom}) we find
$
\dot{\rho}_m+3H\rho_m = \mathcal{I}$,
where
the
interaction function     takes the form
$\mathcal{I}=3\nu H\rho_m$, leading to
\begin{equation}
\rho_m(a)=\rho_{m0}a^{-3(1-\nu)}.
\label{eq:trace_linear_matter_evolution}
\end{equation}
Substitution into the first Friedmann equation
then gives
\begin{equation}
H^2(a)=
\frac{1}{1-\nu}
\left[
c_0+\frac{8\pi G}{3}
\rho_{m0}a^{-3(1-\nu)}
\right].
\label{eq:trace_linear_Ha}
\end{equation}
These expressions coincide exactly with the
matter evolution and expansion history of the
corresponding running vacuum model (see (\ref{eq:rvm_interaction_explicit}) for
$\gamma=0$). Thus, the
torsional reconstruction reproduces simultaneously
the effective dark energy density, its equation
of state, the matter-vacuum interaction, and
both background Friedmann equations.

\begin{figure}[ht!]
\includegraphics[width=0.42\textwidth]{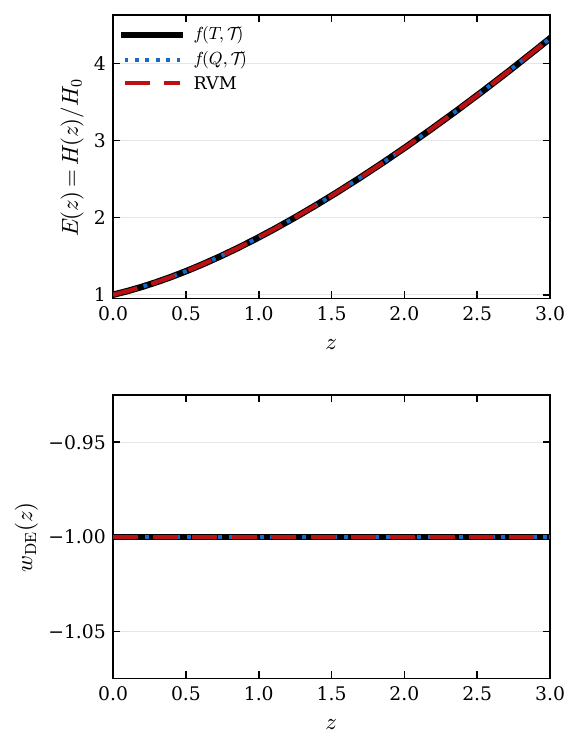}
\caption{{\it{
Comparison of the normalized Hubble expansion
rate $E(z)=H(z)/H_0$ (upper panel) and the
effective dark energy equation-of-state
parameter $w_{\rm DE}(z)$ (lower panel)
in the reconstructed $f(T,\mathcal{T})$
gravity (black solid line),
$f(Q,\mathcal{T})$ gravity with Connection I
(blue dotted line), and running vacuum
cosmology (red dashed line).
The three expansion histories coincide
exactly, and all models satisfy
$w_{\rm DE}=-1$ throughout the displayed
evolution. The parameters are the same
as in Fig.~\ref{quadraticdensity}.
}}}
\label{quadraticevolution}
\end{figure}

We now extend the analysis to
$f(Q,\mathcal{T})$ gravity, adopting Connection I.
Since the two theories possess identical homogeneous
field equations under the common matter prescription,
the corresponding reconstruction follows directly
from the torsional result through $T\rightarrow Q$.
In particular, the gravitational Lagrangian becomes
\begin{equation}
f(Q,\mathcal{T})=
-\frac{2\nu}{1-3\nu}Q
+\frac{16\pi G\nu}{1-3\nu}\mathcal{T}
-\frac{6c_0}{1-3\nu}.
\label{eq:trace_linear_fQ_final}
\end{equation}
For this function, the quantity entering the
second modified Friedmann equation becomes
\begin{equation}
\mathcal{D}_Q=
1+Kf_{\mathcal{T}}
=\frac{1-\nu}{1-3\nu},
\label{eq:trace_linear_Q_condition}
\end{equation}
which satisfies exactly the vacuum condition
(\ref{eq:trace_recon_vacuum_condition}).

Moreover, substituting
Eq.~(\ref{eq:trace_linear_fQ_final})
into the first modified Friedmann equation
gives
\begin{equation}
\rho_m(Q)=
K\left(\frac{1-\nu}{2}Q-3c_0\right),
\label{eq:trace_linear_fQ_matter}
\end{equation}
which is precisely the required running
vacuum matter density. Therefore, the
nonmetricity reconstruction also yields
$\dot H=-4\pi G\rho_m$, together with
$p_{\rm DE}=-\rho_{\rm DE}$ and the
interaction $\mathcal{I}=3\nu H\rho_m$.
Its matter evolution and Hubble expansion
are consequently identical to
Eqs.~(\ref{eq:trace_linear_matter_evolution})
and (\ref{eq:trace_linear_Ha}).

The two reconstructed Lagrangians have identical
functional forms under the correspondence
$T\leftrightarrow Q$ and reproduce exactly
the same RVM background dynamics.
For both theories, regularity of the reconstructed
expressions requires $\nu\neq1/3$, while
$\nu=1$ corresponds to a degenerate branch.
Both theories recover $\Lambda$CDM,
with $\Lambda=3c_0$, as $\nu\rightarrow0$.

This exact correspondence is illustrated in
Figs.~\ref{quadraticdensity} and \ref{quadraticevolution}.
The matter and dark energy density parameters, as well
as the normalized expansion rates, coincide in all three
models, while $w_{\rm DE}=-1$ holds identically.
Thus, the reconstruction reproduces the complete
homogeneous cosmological evolution, including the
matter-vacuum interaction.

Consequently, a separate background observational
analysis would be redundant, since cosmic chronometers,
Type Ia supernovae, and geometrical BAO measurements
yield identical predictions for the same parameters
and calibrations. The observational phenomenology of
RVM and related modified-gravity models has already
been extensively investigated in the literature.

\subsection{Exact reconstruction  of the $
\rho_{\rm DE}(H)=
\frac{3}{8\pi G}
\left(c_0+\nu H^2+\gamma H^4\right)$  running vacuum model}
\label{subsec:rvm_trace_quartic}

We now extend the exact reconstruction to the full running
vacuum density of Eq.~(\ref{eq:trace_recon_requirements}),
with $\gamma\neq0$ and $p_{\rm DE}=-\rho_{\rm DE}$.
Unlike the linear case, this requires nonlinear functions
of the geometrical scalar. We adopt the ansatz
\begin{equation}
f(X,\mathcal{T})=h_X(X)+g_X(X)\mathcal{T},
\label{eq:trace_quartic_ansatz}
\end{equation}
where $X=T$ or $Q$, $\mathcal{T}=-\rho_m$, and the
required matter density $\rho_m(X)$ is given by
Eq.~(\ref{eq:trace_recon_matter}).

Substituting this ansatz into the first modified
Friedmann equation and the vacuum condition
(\ref{eq:trace_recon_vacuum_condition}), we obtain
\begin{equation}
h_X-2Xh_X'={}-X+\frac{2\rho_m}{K}
+3\rho_mg_X -2X\rho_mg_X',
\label{eq:trace_quartic_h_equation}
\end{equation}
\begin{equation}
2\rho_mg_X'
+\left[3\rho_m'-K\right]g_X
 = \left(1-\frac{2\rho_m'}{K}\right),
\label{eq:trace_quartic_g_equation}
\end{equation}
where primes denote derivatives with respect to $X$,
and $K=(8\pi G)^{-1}$. Thus, the reconstruction equations are identical
for torsion and nonmetricity under $T\leftrightarrow Q$.
The second equation determines $g_X$
through a first-order linear differential equation,
while the first subsequently determines $h_X$.
Hence, local reconstruction is possible on regular
intervals with $\rho_m\neq0$, without imposing any
special relation between $c_0$, $\nu$, and $\gamma$.

To identify a physically admissible branch, we introduce
$y=X/(6H_0^2)=H^2/H_0^2$,
$C=c_0/H_0^2$, and $\Gamma=\gamma H_0^2$. Then
\begin{equation}
\rho_m(y)=3KH_0^2F(y),\qquad
F(y)=(1-\nu)y-C-\Gamma y^2,
\label{eq:trace_quartic_F}
\end{equation}
with $C=1-\Omega_{m0}-\nu-\Gamma$.
For $C>0$, $\nu\geq0$, and $\Gamma>0$, the
effective vacuum density is positive. Defining
\begin{equation}
d=\sqrt{(1-\nu)^2-4\Gamma C},\qquad
y_\pm=\frac{1-\nu\pm d}{2\Gamma},
\label{eq:trace_quartic_roots}
\end{equation}
we have $F(y)=\Gamma(y-y_-)(y_+-y)$.
Thus, for $d>0$ and $0<y_-<1<y_+$, both matter
and vacuum densities remain positive on the interval
$y_-<y<y_+$, with $y_-$ corresponding to the future
de Sitter attractor.

Writing $g_X(X)=U(y)/K$, where $U(y)$ is common
to both gravitational formulations, the reconstruction
equation becomes
\begin{equation}
F\frac{dU}{dy}
+\left(\frac{3}{2}\frac{dF}{dy}-1\right)U
=1-\frac{dF}{dy}.
\label{eq:trace_quartic_U_ode}
\end{equation}
For the parameter branch $d>2/3$, we introduce
$t=(y-y_-)/(y_+-y_-)$,
$A=3/2-1/d$, and $B=3/2+1/d$.
The solution regular at the future de Sitter
endpoint is then
\begin{equation}
U(t)=
\frac{1}{d\,t^A(1-t)^B}
\bigg[
(1-d)B_t(A,B)
+2d\,B_t(A+1,B)
\bigg],
\label{eq:trace_quartic_U_solution}
\end{equation}
where $B_t(a,b)$ denotes the incomplete beta function,
which can equivalently be expressed through the
Gauss hypergeometric function as
\begin{equation}
B_t(a,b)=\frac{t^a}{a}
{}_2F_1(a,1-b;a+1;t).
\label{eq:trace_quartic_beta}
\end{equation}

\begin{figure}[ht!]
\includegraphics[width=0.42\textwidth]{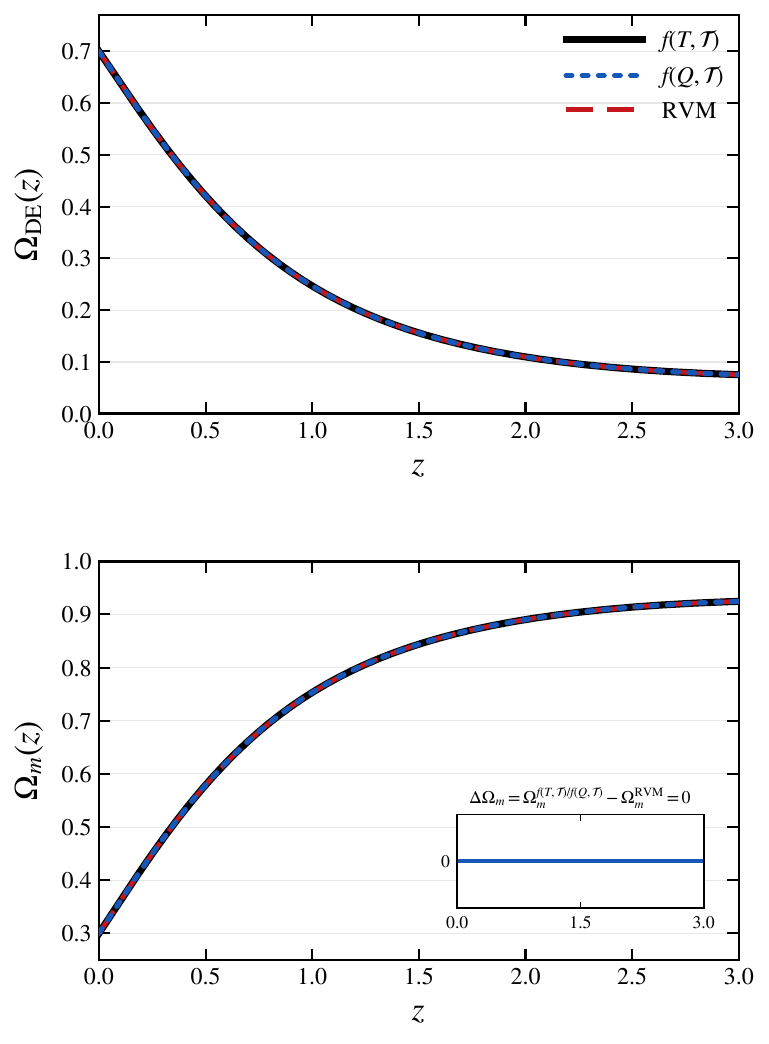}
\caption{{\it{Dark energy density parameter $\Omega_{\rm DE}(z)$
(upper panel) and matter density parameter $\Omega_m(z)$
(lower panel) in the positive-$\gamma$ quartic reconstruction.
The black solid, blue dotted, and red dashed curves
represent $f(T,\mathcal T)$, $f(Q,\mathcal T)$ (Connection I),
and running vacuum cosmology, respectively.
The three models coincide exactly. The inset shows
the vanishing difference in $\Omega_m$ between
either reconstructed model and RVM.
We use
$\Omega_{m0}=0.30$, $\nu=0.02$,
$\gamma H_0^2=10^{-3}$, and $c_0/H_0^2=0.679$.}}}
\label{fig:quartic_density_final}
\end{figure}

The general solution also contains a homogeneous term
proportional to $t^{-A}(1-t)^{-B}$.
On the considered branch, this contribution diverges
as $t\rightarrow0$. Requiring regularity at the future
de Sitter endpoint therefore eliminates this term,
yielding
\begin{equation}
U(y_-)=\frac{2(1-d)}{3d-2}.
\label{eq:trace_quartic_endpoint}
\end{equation}
In the limit $\Gamma\rightarrow0$, the solution
recovers the common linear reconstruction of
Sec.~\ref{subsec:rvm_trace_linear}, namely
$U=2\nu/(1-3\nu)$.

The remaining geometrical function follows from
a single quadrature. Defining
$h_X(X)=H_0^2\mathfrak h(y)$, we obtain
\begin{equation}
\mathfrak h(y)=\sqrt y\left[
C_h-\frac12\int_{y_-}^y
\frac{\mathcal S(s)}{s^{3/2}}\,ds
\right],
\label{eq:trace_quartic_h_solution}
\end{equation}
where
\begin{equation}
\mathcal S(y)=-6y+6F(y)+9F(y)U(y)
 -6yF(y)\frac{dU}{dy}.
\label{eq:trace_quartic_S}
\end{equation}
Consequently, the reconstructed gravitational
Lagrangians are
\begin{align}
f(T,\mathcal T)={}&H_0^2\mathfrak h
\left(\frac{T}{6H_0^2}\right)
+\frac{\mathcal T}{K}U
\left(\frac{T}{6H_0^2}\right),
\label{eq:trace_quartic_fT}\\
f(Q,\mathcal T)={}&H_0^2\mathfrak h
\left(\frac{Q}{6H_0^2}\right)
+\frac{\mathcal T}{K}U
\left(\frac{Q}{6H_0^2}\right).
\label{eq:trace_quartic_fQ}
\end{align}
The integration constant $C_h$ represents the
homogeneous $\sqrt X$ freedom, which does not
affect the background equations. We choose the
same value of this constant in the representative
torsional and nonmetricity reconstructions, although
independent choices would generate different
Lagrangians with identical homogeneous cosmological
evolution. Thus, regularity determines the common
trace-coupling function within the considered
ansatz and parameter branch, while the complete
Lagrangian is not uniquely determined.

\begin{figure}[ht!]
\includegraphics[width=0.42\textwidth]{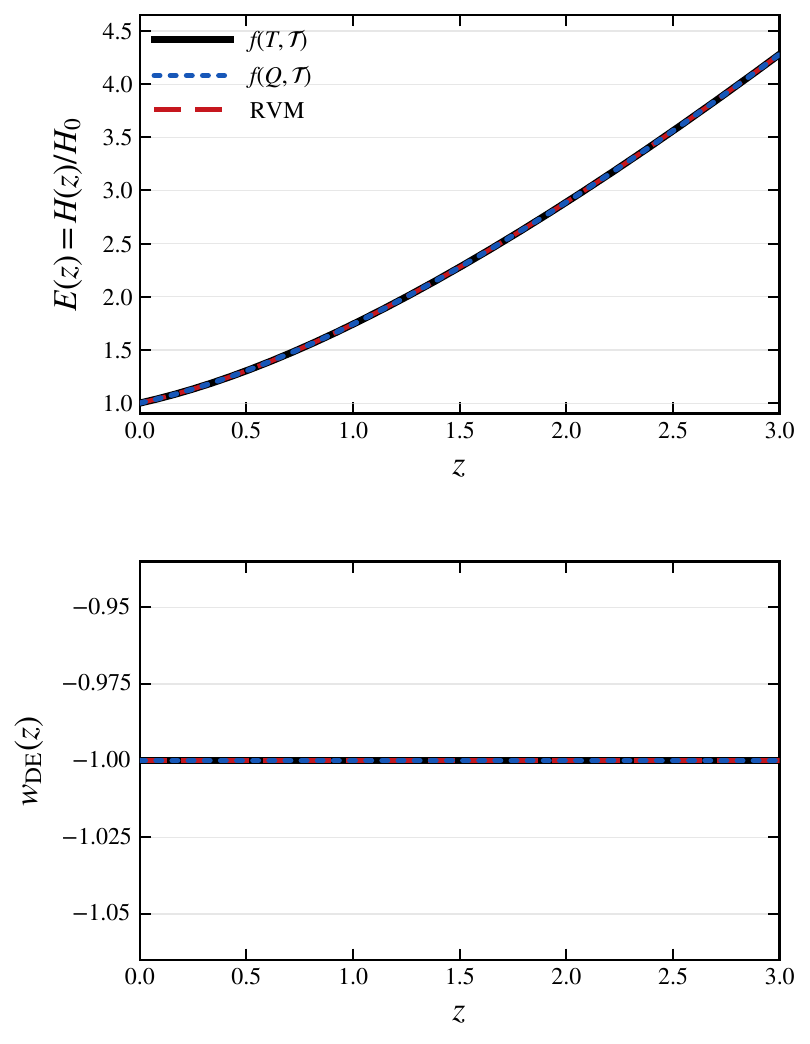}
\caption{{\it{Normalized expansion rate $E(z)=H(z)/H_0$
(upper panel) and effective dark energy equation of
state $w_{\rm DE}(z)$ (lower panel) in the positive-$\gamma$
quartic reconstruction. The black solid, blue dotted,
and red dashed curves correspond to $f(T,\mathcal T)$,
$f(Q,\mathcal T)$, and running vacuum cosmology,
respectively. Their expansion histories coincide and
the strict vacuum equation of state holds throughout
the displayed interval. The parameters are the same
as in Fig.~\ref{fig:quartic_density_final}.}}}
\label{fig:quartic_expansion_final}
\end{figure}

\begin{table*}[ht!]
\caption{Comparison of the different levels of
correspondence between running vacuum cosmology and
the gravitational theories considered in this work.
The last column refers to equivalence at the
homogeneous cosmological level. For the nonminimal
geometry-matter coupling theories, the conclusions
refer to the matter prescription adopted in
Sec.~\ref{subsec:rvm_nonminimal}.
The nonlinear trace-coupled reconstruction is
established on the cosmological branch discussed
in Sec.~\ref{subsec:rvm_trace_quartic}.}
\label{tab:rvm_correspondence}
\begin{ruledtabular}
\begin{tabular}{lccc}
Gravitational theory
& Running vacuum density
& $w_{\rm DE}=-1$
& Full RVM background equivalence \\
\hline
$f(T)$ and $f(Q)$
& Yes & Generally no & No \\
Nonminimal geometry-matter coupling
& Yes & Only for constant $\rho_{\rm DE}$
& Only $\Lambda$CDM \\
Trace coupling ($c_0+\nu H^2$)
& Yes & Yes & Yes \\
Trace coupling ($c_0+\nu H^2+\gamma H^4$)
& Yes & Yes & Yes \\
\end{tabular}
\end{ruledtabular}
\end{table*}

The exact correspondence can be verified directly.
In particular, the acceleration denominators satisfy
$\mathcal D_T=\mathcal D_Q=1+U(y)$,
precisely reproducing the vacuum condition
(\ref{eq:trace_recon_vacuum_condition}).
On regular branches, both Friedmann equations yield
$\dot H=-4\pi G\rho_m$ and
\begin{eqnarray}
&&\rho_{\rm DE}(H)=3KH_0^2
\left(C+\nu y+\Gamma y^2\right),\\
&& p_{\rm DE}=-\rho_{\rm DE},
\label{eq:trace_quartic_density}
\end{eqnarray}
together with the matter-vacuum interaction
$\mathcal I=3H(\nu+2\Gamma y)\rho_m$.

For the $f(Q,\mathcal T)$ realization with
Connection I, the independent connection equation
is also satisfied. Indeed, for vanishing hypermomentum,
the connection equation takes the form
$
\nabla_\mu\nabla_\nu\left[
\sqrt{-g}\,(-1+f_Q)P_\alpha{}^{\mu\nu}
\right]=0$.
In the coincident FLRW gauge, all background
quantities depend only on cosmic time, while
$P_\alpha{}^{00}=0$. Consequently, the
connection equation is identically satisfied.

The common expansion history follows from
$dy/dz=3F(y)/(1+z)$, with $y(0)=1$, yielding
\begin{equation}
E^2(z)=\frac{y_-+y_+R(z)}{1+R(z)},\qquad
R(z)=\frac{1-y_-}{y_+-1}(1+z)^{3d}.
\label{eq:trace_quartic_Ez}
\end{equation}
Therefore, the reconstructed $f(T,\mathcal T)$
and $f(Q,\mathcal T)$ theories reproduce exactly
the running vacuum expansion history, matter
evolution, and effective vacuum density.

As an explicit example, we choose
$\Omega_{m0}=0.30$, $\nu=0.02$,
$\Gamma=10^{-3}$, and $C=0.679$, obtaining
$d\simeq0.9786$, $y_-\simeq0.6933$, and
$y_+\simeq979.31$. The common regular solution
has the finite endpoint value
$U(y_-)\simeq0.04571$.
Furthermore, throughout $0\leq z\leq3$,
both density parameters remain positive, while
$1.0462\lesssim\mathcal D_T=\mathcal D_Q
\lesssim1.0726$.
Thus, the acceleration equations remain regular
over the considered interval.

The resulting cosmological evolution is presented
in Figs.~\ref{fig:quartic_density_final}
and \ref{fig:quartic_expansion_final}.
The torsional and nonmetricity reconstructions
coincide exactly with the RVM scenario in their matter
and dark energy density parameters, normalized Hubble
expansion, and strict equation of state
$w_{\rm DE}=-1$. Hence, the inclusion of a genuine
$H^4$ contribution does not obstruct complete
running vacuum reconstruction at the homogeneous
cosmological level.

The above reconstruction establishes the existence
of regular, positive-density solutions extending from
the future de Sitter endpoint through $z=3$ within
the adopted homogeneous dust framework.
We emphasize that regularity is imposed at the
future de Sitter endpoint $y_-$.
For the illustrative positive-$\Gamma$ branch,
the common reconstructed trace-coupling function
becomes singular as $y\rightarrow y_+$, and hence
the upper root does not represent a second
regular endpoint of the same reconstruction.
The present dust reconstruction does not provide
a complete early-Universe description, which would
require including radiation. Furthermore, exact
background equivalence does not guarantee identical
structure formation, whose investigation lies
beyond the scope of the present work.

\subsection{Density matching versus full cosmological equivalence}
\label{subsec:density_vs_equivalence}

Our analysis demonstrates that reproducing the running
vacuum density $\rho_{\rm DE}(H)$ does not necessarily
imply equivalence with running vacuum cosmology.
As shown in Sec.~\ref{subsec:rvm_fT_fQ}, minimally coupled
$f(T)$ and $f(Q)$ gravity reproduce the desired energy
density but generally yield different equations of state
and expansion histories. Similarly, for the nonminimal
geometry-matter couplings of Sec.~\ref{subsec:rvm_nonminimal},
within the adopted matter prescription, imposing
$w_{\rm DE}=-1$ necessarily leads to a constant
effective dark energy density. In contrast,
trace-coupled gravity allows the simultaneous
reconstruction of $\rho_{\rm DE}(H)$ and the strict
vacuum equation of state, recovering the corresponding
matter-vacuum interaction and complete homogeneous
cosmological evolution.

Our main findings are summarized in
Table~\ref{tab:rvm_correspondence}. The linear
trace-coupled reconstructions reproduce exactly
the constant-plus-$H^2$ running vacuum scenario,
while the nonlinear constructions extend this
equivalence to include the $H^4$ contribution,
without requiring special relations between the
running vacuum parameters. In both cases, the
torsional and nonmetricity reconstructions have
identical functional forms under $T\leftrightarrow Q$,
provided the same perfect-fluid matter prescription
is adopted. Moreover, the nonlinear reconstruction
admits regular cosmological branches with positive
matter and effective vacuum densities.

An important consequence is that theories reproducing
the same $H(z)$ cannot be distinguished through
background geometrical observations, such as cosmic
chronometers, Type Ia supernovae, and BAO measurements,
provided the relevant calibrations are consistently
matched. Nevertheless, identical expansion histories
and functional forms of the reconstructed Lagrangians
do not imply complete physical equivalence.
The distinct geometrical structures of torsion and
nonmetricity gravity may lead to different cosmological
perturbations, effective gravitational couplings,
and structure formation histories.

Finally, the positive-$\gamma$ reconstruction of
Sec.~\ref{subsec:rvm_trace_quartic} provides explicit
regular solutions with positive matter and vacuum
densities throughout $0\leq z\leq3$, approaching
a regular future de Sitter attractor. However,
these results concern the homogeneous dust
background and do not establish complete
phenomenological viability. Extending the analysis
to include radiation, examining cosmological
perturbations, and investigating the associated
structure-growth predictions remain important
directions for future work.

Hence, trace-coupled torsion and nonmetricity gravity
provide explicit realizations of the complete running
vacuum background dynamics. In particular, the same
functional reconstruction can be implemented in both
geometrical frameworks, establishing a correspondence
that extends beyond the mere matching of the effective
dark energy density, while preserving the distinction
between background equivalence and full gravitational
equivalence.

\section{Conclusions}
\label{sec:conclusions}

Torsional and nonmetricity formulations of gravity provide
alternative geometrical frameworks for describing the
gravitational interaction, allowing for a variety of
extensions beyond general relativity. In particular,
$f(T)$ and $f(Q)$ gravity, together with their
nonminimal matter-coupled generalizations, offer
interesting possibilities for explaining the late-time
acceleration of the Universe. On the other hand,
running vacuum cosmology provides a different approach,
in which the vacuum energy density depends on the
Hubble parameter while retaining the strict equation
of state $w_{\rm DE}=-1$. For constant Newton's
gravitational coupling, this evolution necessarily
induces an energy exchange between matter and vacuum,
making RVM an intrinsically interacting dark energy
scenario.

A central theoretical question is whether this
phenomenological running vacuum description can
emerge from an underlying gravitational Lagrangian.
Although quantum field theory in curved spacetime
provides motivations for a dynamical vacuum density,
the existence of a gravitational action reproducing
the complete running vacuum evolution is not
automatically guaranteed. In particular, matching
the characteristic dependence
$\rho_{\rm DE}(H)\propto c_0+\nu H^2+\gamma H^4$
does not ensure that the corresponding gravitational
theory also reproduces the strict vacuum equation
of state and the associated matter-vacuum interaction.
Motivated by this distinction, in the present work
we investigated the reconstruction of running vacuum
cosmology within torsion and nonmetricity gravity,
considering minimally coupled, nonminimal
geometry-matter-coupled, and trace-coupled theories.

We first demonstrated that the running vacuum density
can be reconstructed analytically in minimally coupled
$f(T)$ and $f(Q)$ gravity. Nevertheless, the resulting
effective dark energy generally possesses an evolving
equation of state, while the matter conservation law
and expansion history differ from those of RVM.
We quantified these differences using Pantheon+
supernovae, cosmic chronometers, and DESI DR2 BAO
measurements. Although the geometrical reconstructions
and RVM lead to different parameter constraints and
statistical fits, the information criteria do not
provide a decisive preference between them.
We further examined nonminimal geometry-matter
couplings and showed that, within the adopted matter
prescription, imposing the strict vacuum equation of
state restricts the effective dark energy density
to a constant, preventing the realization of a
genuinely evolving running vacuum.

Our main result concerns trace-coupled
$f(T,\mathcal{T})$ and $f(Q,\mathcal{T})$ gravity,
for which we obtained explicit gravitational
Lagrangians reproducing the complete RVM dynamics
at the homogeneous cosmological level.
For the constant and $H^2$ terms, this correspondence
is achieved through simple linear trace couplings,
while the inclusion of a genuine $H^4$ contribution
requires nonlinear gravitational functions, whose
analytical reconstruction involves incomplete beta
and hypergeometric functions. Importantly, adopting
a common perfect-fluid matter variation prescription,
we showed that the torsional and nonmetricity
reconstructions possess identical functional forms
under $T\leftrightarrow Q$. Furthermore, we identified
regular solutions approaching a future de Sitter
attractor, including positive-$\gamma$ branches
with positive matter and effective vacuum densities.
The reconstructed theories satisfy both modified
Friedmann equations and reproduce simultaneously
the prescribed vacuum density, $w_{\rm DE}=-1$,
the characteristic matter-vacuum interaction,
and the exact RVM expansion history.
Thus, our results establish that running vacuum
cosmology can be realized directly through modified
gravitational dynamics, rather than merely being
introduced as an effective interacting vacuum
component. Moreover, they demonstrate that the
same functional gravitational reconstruction can
be implemented in two distinct geometrical
frameworks, leading to identical homogeneous
cosmological evolution without implying complete
gravitational equivalence.

Several directions deserve further investigation.
A natural extension is to include radiation and examine
the cosmological evolution of the reconstructed theories
in the early Universe, particularly in the presence
of the $H^4$ contribution. More importantly, our
results provide a new perspective on the perturbative
description of running vacuum cosmology. While the
conventional RVM framework specifies the homogeneous
evolution, its perturbative implementation requires
additional assumptions concerning vacuum fluctuations
and matter-vacuum energy-momentum transfer.
Our reconstructed gravitational Lagrangians offer
a possible action-based realization of RVM, providing
a framework in which cosmological perturbations can
be investigated from the underlying gravitational
dynamics. The investigation of scalar, vector, and
tensor perturbations, together with the identification
of propagating degrees of freedom and the examination
of ghost and gradient stability, will therefore be
particularly important. The role of the matter action
and the independent connection also deserves further
study. Moreover, combining background observations
with structure-growth and gravitational perturbation
measurements could provide additional constraints on
the reconstructed theories. In this respect, our
results open a promising direction toward connecting
the phenomenological running vacuum scenario with
a consistent gravitational description, potentially
extending its predictive framework beyond the
homogeneous cosmological evolution.

\begin{acknowledgments}  The authors acknowledge the contributions of the LISA
CosWG  and  COST
Actions
 CA21106 ``COSMIC WISPers
in the Dark Universe: Theory, astrophysics and experiments'',  CA21136
``Addressing observational tensions in cosmology with
  systematics and fundamental physics (CosmoVerse)'',    CA23130
``Bridging high and low energies in
search of quantum gravity (BridgeQG)'', and CA24101 ``Testing Fundamental
Physics with Seismology''.
 \end{acknowledgments}
 
 
\end{document}